\documentclass[a4paper,11pt]{article}
\pdfoutput=1 

\usepackage{jcappub} 

\usepackage{cases}
\usepackage{amsmath}
\usepackage{amsfonts}
\usepackage{amssymb}
\usepackage{dcolumn}
\usepackage{bm}
\usepackage{bbm}
\usepackage{xcolor}
\usepackage{array}
\usepackage{subfigure}
\usepackage{wasysym}
\usepackage{mathtools}
\usepackage{hyperref}
\usepackage{graphicx} 
\usepackage{tikz}

\usepackage{appendix}

\hypersetup{colorlinks=true,
	breaklinks=true,
	pdfstartview=Fit,
	linkcolor=blue,
	citecolor=blue,
	urlcolor=blue
}
\usepackage{cleveref}

\definecolor{myblue}{RGB}{0, 100, 200}
\definecolor{myred}{RGB}{214, 39, 40}
\usetikzlibrary{
    arrows.meta,
    positioning,
    fit,
    backgrounds,
    calc
}

\definecolor{forwardblue}{HTML}{EAF2F8}
\definecolor{forwardline}{HTML}{315F7D}
\definecolor{inversegreen}{HTML}{EAF6F1}
\definecolor{inverseline}{HTML}{397565}
\definecolor{textdark}{HTML}{263238}

\title{\boldmath Reconstructing the Dark Matter Equation of State with Compact Object Inspirals}

\author[a]{Boris Betancourt Kamenetskaia,}
\author[a]{Qianhang Ding,}
\author[a]{Hui-Yu Zhu}

\affiliation[a]{Cosmology, Gravity and Astroparticle Physics Group, Center for Theoretical Physics of the Universe, Institute for Basic Science (IBS), Daejeon, 34126, Korea}

\emailAdd{laybors@ibs.re.kr}
\emailAdd{dingqh@ibs.re.kr}
\emailAdd{hzhuav@ibs.re.kr}

\abstract{We investigate the gravitational wave (GW) signatures of compact binaries embedded in extended dark matter (DM) configurations in hydrostatic equilibrium. If a neutron star is surrounded by a sufficiently massive and extended envelope, an inspiraling companion experiences dynamical friction (DF) in addition to the standard GW energy loss. We show that this environmental effect can be isolated through a GW observable, the $D$-function, which directly characterizes the additional dissipative power induced by the surrounding medium. Since the density of the envelope is directly determined by the DM equation of state together with the stellar boundary conditions, the frequency dependence of the $D$-function provides direct information about the macroscopic properties of the dark sector. We develop a regression-based framework to reconstruct both the density profile and the DM equation of state over the density range probed by observations of the GW inspiral. Using representative DM models as benchmarks, we demonstrate that the method accurately recovers the input equation of state while remaining largely independent of the microscopic realization of DM. Our results show that future GW observations of compact binaries can provide a probe of the equation of state of DM.}

\begin{document}
\maketitle
\flushbottom

\section{Introduction } 

Dark matter (DM) constitutes approximately $27\%$ of the energy density of the Universe~\cite{Planck:2018vyg}, yet its fundamental nature remains one of the central open questions in modern physics. Although its existence is firmly established through observations ranging from galactic rotation curves to cosmological measurements, it remains unknown whether DM consists of elementary particles, compact astrophysical objects, or some entirely different form of matter (for reviews on different DM models, see \cite{Bertone:2004pz,Feng:2010gw, SciPostPhysRev.1}). Among the various possibilities, particle DM remains one of the most extensively studied frameworks owing to its strong theoretical motivation and its ability to account for the observed cosmological and astrophysical properties of DM.

A particularly fascinating consequence of the particle DM paradigm is the possibility that DM can form gravitationally bound, macroscopic configurations if its microscopic properties give rise to an effective pressure capable of counteracting gravity. Such pressure can originate from quantum effects, self-interactions, or a combination of both, depending on the underlying particle model. This idea dates back to Kaup~\cite{Kaup:1968zz}, who studied a self-gravitating system of complex scalar fields, and has since been extended to a wide variety of particle models: bosonic DM with repulsive or attractive self‑interactions \cite{Colpi:1986ye,Eby:2015hsq}, fermionic DM supported by degeneracy pressure or self‑interactions \cite{Kouvaris:2015rea}, and many others. The equilibrium structure, stability, and dynamics of such self-gravitating
compact configurations have been studied extensively for both bosonic and
fermionic matter~\cite{RuffiniBonazzola1969,Jetzer1992BosonStars,
SchunckMielke2003BosonStars,LieblingPalenzuela2023,
Narain2006FermionStars}.\footnote{For broader reviews of boson stars and other
horizonless compact objects, see
Refs.~\cite{LieblingPalenzuela2023,CardosoPani2019}.}
Within this paper we refer to such self-gravitating, macroscopically extended configurations composed predominantly of DM as dark stars (DSs). Regardless of the underlying particle physics, they are described by hydrostatic equilibrium, with their masses, radii, and internal density profiles determined by the DM equation of state (EoS). Depending on the DM particle mass and interaction strength, DSs can span many orders of magnitude in size and mass, ranging from subsolar masses to above $10^{5}\,M_\odot$, with radii extending from compact stellar scales to several solar radii or more, while remaining stable over cosmological timescales.

Several observational techniques constrain the presence of compact dark objects through their gravitational interactions. These include, for example, stellar microlensing~ \cite{Macho:2000nvd,EROS-2:2006ryy,Niikura:2019kqi}, supernova magnification~\cite{Zumalacarregui:2017qqd}, gravitational wave searches~\cite{LVK:2022ydq,Maselli2017DarkStars,Sennett2017BosonStarTides,
Pacilio2020ParticlePhysics,Wystub2023DarkStars,
Celato2025Fmode}, and the dynamical heating of dwarf galaxies~\cite{Brandt:2016aco}. Although originally developed to search for compact DM candidates such as Massive Compact Halo Objects (MACHOs) and primordial black holes (PBHs), these observations can also be interpreted as limits on compact DSs, implying that they can constitute at most a small fraction of the total DM population, namely a few percent~\cite{Green:2020jor,SciPostPhysRev.1}. While these observations constrain the abundance, masses, and compactness of DSs, they provide only limited information about their internal structure and, in particular, the underlying DM EoS. Developing observational probes that are directly sensitive to the density profile of extended DM configurations would therefore open a complementary window onto the macroscopic properties of particle DM and, in particular, its EoS.


When we consider a DS in a binary system, the binary separation usually decreases due to gravitational wave (GW) emission~\cite{Peters:1964zz}. However, if one of the compact objects is surrounded by a dense dark environment, and if the companion moves inside this environment, there will be an additional power loss due to dynamical friction (DF), which is a dissipative phenomenon experienced by a massive object moving through a medium~\cite{1943ApJ....97..255C}. More specifically, during the inspiral, a wake forms behind the moving object, and the gravitational interaction between the object and this wake induces an additional drag force, leading to extra power loss~\cite{Baumann:2021fkf,Tomaselli:2023ysb,Dosopoulou:2023umg,Ding:2025nxe}. This effect has been widely studied since the seminal work of Chandrasekhar~\cite{1943ApJ....97..255C}, and has been generalized to various types of media, including collisional gas~\cite{Ostriker:1998fa}, superfluids~\cite{Berezhiani:2019pzd}, and fuzzy DM~\cite{Lancaster:2019mde}.

Since DF originates from the backreaction associated with the exchange of energy and angular momentum between the companion and its surrounding environment, the resulting additional energy loss encodes information about the properties of the environment, such as its density profile and EoS. Over a sufficiently long observation period, this additional dissipation can accumulate into a measurable GW dephasing, making GWs a promising probe of the properties of DM environments~\cite{Eda:2014kra,Chen:2020lpq,Coogan:2021uqv,Li:2021pxf,Bertone:2019irm,Berezhiani:2023vlo,Takatsy:2025bfk,Ding:2025nxe,Ding:2025hqf,BarausseCardosoPani2014,Kavanagh2020DMDynamics,
Cole2023EnvironmentalEffects}. In a previous study, we constructed a quantity, denoted by $D$, that depends only on GW observables and characterizes the waveform dephasing induced by the additional power loss in binary black hole systems~\cite{Ding:2025hqf}. This quantity provides a way to infer the presence of a dark environment around the black hole and to reconstruct its density profile, which helps us characterize the DM properties. In this work, we generalize this idea to a DS with a neutron star (NS) at its center, which has been widely studied within relativistic two-fluid and admixed-star formalisms~\cite{SandinCiarcelluti2009,LeungChuLin2011,Kain2021DANS,
DasMalikNayak2022,Diedrichs:2023trk}. We aim to extract further properties of the DS, such as its equation of state. Specifically, we consider an inspiraling binary consisting of a NS embedded at the center of a DS and a compact companion (either a NS or a black hole) moving through the surrounding DM envelope.

While the density profile characterizes an individual star, the EoS is a universal property of the underlying DM. More importantly, the EoS provides the macroscopic behavior of the microphysics in the dark sector. Distinct DM models, particle masses and interaction strengths lead to different pressure-density relations in general and, thus, to different equilibrium configurations of DSs. Determining the DM EoS would therefore provide valuable information about the fundamental properties of particle DM, in the same way that measurements of the NS EoS constrain the physics of dense nuclear matter~\cite{Hinderer2008TidalLove,Read2009ParameterizedEOS,
LindblomIndik2012Inverse,LindblomIndik2014InverseII,
Abbott2018GW170817EOS}. 

In this work, we use a regression method to infer the EoS and density profile of the DS surrounding a NS, based on the frequency dependence of the GW observable $D$ in a neutron-star binary system. As examples, we consider both boson star and fermion star environments. For each case, we generate mock data from a reference EoS and show that the regression successfully recovers the input model with high accuracy. Since the EoS (with corresponding central and boundary conditions) uniquely determines the equilibrium density profile of the DS, this procedure allows us to reconstruct both quantities from the same GW observable. We also identify the region of parameter space in which the effect is sufficiently large to be observed, and find that it lies within the sensitivity band of Einstein Telescope \cite{Punturo:2010zz} and DECIGO \cite{Kawamura:2006up}. Throughout this study, we restrict our analysis to circular and coplanar orbits for simplicity.

The paper is organized as follows: In Sec.~\ref{Sect.DarkStar}, we briefly describe how a DS-NS system forms and calculate the resulting density profile of the DS after the formation of the DS-NS system. In Sec.~\ref{Sect.DynamicalFriction}, we discuss how DF induces GW dephasing and how the properties of the dense dark environment can be related to the GW observable $D$. In Sec.~\ref{sec:examples_DS}, we discuss the analytical and numerical properties of the benchmark DSs. In Sec.~\ref{Sect.Cloud_Properties}, we explain how the EoS and density profile of the DS can be inferred using a regression-based method, and quantify the corresponding fitting errors. In Sec.~\ref{Sect.ObservationalConstraints}, we derive several observational constraints relevant to this effect. Finally, we summarize our results and discuss future directions in Sec.~\ref{Sect.DiscussionandConclusion}.

\section{Dark Star--Neutron Star System} 
\label{Sect.DarkStar}

The DS-NS system refers to a NS as the dense core enveloped by an extended DS, 
which is in two-fluid hydrostatic equilibrium. This composite astrophysical system offers a unique laboratory to study the nature of DSs, where the NS serves as an electromagnetic probe and potentially hints at the existence  
of the DS \cite{Diedrichs:2023trk}.  In particular, introducing a compact binary component, such as a 
NS, into this DS-NS system would lead to the GW 
signal which contains the information of the orbital change of the binary system. 
This offers a special channel to probe the structure of the surrounding DS by studying the corresponding DF signal in GWs \cite{Ding:2025hqf}. In this section, we first discuss the formation mechanism of the DS-NS system, and then define the static two-fluid background structure of the DS–NS system. 

\subsection{Formation mechanism} 

Several mechanisms have been proposed for the formation of a DS--NS system:
(i) NS formation inside a newly formed DS \cite{Alvarez-Rios:2024kfr, Leung:2019ctw, Chan:2023atg};
(ii) dynamical capture and merger between a NS and a DS \cite{Dietrich:2018bvi}; and
(iii) accretion of DM onto a NS \cite{DiGiovanni:2020frc, Brito:2015yfh}. These formation mechanisms provide the theoretical foundation for the potential existence of the DS-NS system. At present, however, there is no conclusive observational evidence for their existence. Some unusual compact-object systems, such as XTE J1814-338 and HESS J1731-347, have been proposed as possible DS-NS candidates \cite{Pitz:2024xvh, Yang:2024ycl, Lopes:2024ixl}, although further evidence is required to distinguish this interpretation from alternative possibilities, such as a soft EoS or a strange star \cite{Sagun:2023rzp}. Once a compact object begins to inspiraling inside a DS-NS system, it will influence the GW signal because of DF. This imprint could provide evidence for the existence of a DS-NS system while also offering a means of reconstructing the EoS of the DM.

In the following, we first discuss several possible channels for forming a DS-NS composite system. These include the three formation channels of a NS inside a pre-existing DS we mentioned before:

\begin{itemize}

\item \textbf{Neutron star formation inside a newly formed dark star:}

In an overdense region, DM and baryonic matter can become gravitationally bound and accumulate within the same potential well \cite{Alvarez-Rios:2024kfr}. 
First, the DM can form a self-gravitating structure through dissipative or gravitational cooling. For example, scalar-wave emission can gravitationally cool bosonic DM and lead to the formation of a bosonic DS \cite{Seidel:1993zk}, while a fermionic DS can form through radiation from the dark sector \cite{Fan:2013yva,Chang:2018bgx}. Then, as the extended DS forms, its gravitational potential draws surrounding baryonic gas toward the center, increasing the concentration of baryonic matter.

As for the star formation inside the DS, the accumulated baryonic gas must undergo further cooling and transfer angular momentum to its surroundings.
Thus, the cooling timescale should be shorter than the free-fall timescale, such that $t_{\rm cool}<t_{\rm ff}$. The resulting high-density baryonic core can subsequently undergo stellar evolution, eventually forming a massive star or a massive white-dwarf core. Once electron degeneracy pressure is too small to support the core against gravitational collapse, the resulting massive stellar core will collapse to form a NS at the center of the DS \cite{Leung:2019ctw, Chan:2023atg}.

\item \textbf{Dynamical capture and merger between a neutron star and a dark star:}

When a NS encounters a DS, a dissipative mechanism is required to form a gravitationally bound DS--NS system, 
otherwise a conservative two-body encounter results only in a flyby. The energy dissipated during a close encounter, $\Delta E$, must satisfy $\Delta E > \frac{1}{2}\mu v_{\infty}^2$, where $\mu\equiv M_{\rm DS} M_{\rm NS}/(M_{\rm DS}+ M_{\rm NS})$ is the reduced mass of the NS-DS system and $v_{\infty}$ is their relative velocity at infinity. Several mechanisms can cause the required energy loss, including GW emission during a relativistic close encounter \cite{Hansen:1972jt}, DF power loss \cite{Macedo:2013qea, Lancaster:2019mde}, and energy exchange during three-body interactions \cite{1976A&A....53..259A, Ginat:2024npu}. Once a bound DS-NS system forms, continued GW emission and DF drive the subsequent inspiral and merger \cite{Dietrich:2018bvi}. The resulting configuration is a composite object containing a NS embedded at the center of the DS. However, dynamical capture between a NS and a DS is expected to be rare in typical cosmic environments because of the small spatial overlap between the objects and the low number density of compact objects, while the dense stellar environments, such as globular clusters and nuclear star clusters, can substantially enhance the capture probability by increasing the local population density of compact objects \cite{Ivanova:2007bu}. 

\vspace{0.5cm}
\item \textbf{Accretion of dark matter on a neutron star:}

A NS embedded in a DM environment provides a deep gravitational potential that can capture DM particles. As DM particles fall toward the NS, they can transfer kinetic energy to the stellar medium through dissipative processes such as scattering with nucleons or interactions with dark radiation. After repeated scattering events, the DM particles lose orbital energy and become concentrated toward the center of the NS, forming a DM core \cite{Bell:2020jou}. However, to form an extended DM envelope around the NS, an additional self-gravitating DM component is required, such as a bound scalar cloud. The gravitational potential of the NS can then induce further contraction of the self-gravitating scalar field, while scalar-wave emission provides a channel for gravitational cooling and allows a bound fermion-boson configuration to form \cite{Seidel:1993zk, DiGiovanni:2020frc}.
\end{itemize}

The formation channels discussed above produce a DS-NS system containing a single NS. However, we need an additional compact object as for the GW detection, and therefore a subsequent mechanism is needed to form a binary NS embedded within the DS. There are two channels to form such a structure.
\begin{itemize}
\item In the first channel, 
the DM and the baryonic matter evolve together within the same gravitational potential well. The baryonic gas cools and fragments to form massive stars, while the DM forms a DS through gravitational condensation or dark-sector cooling \cite{Mocz:2019pyf, Veltmaat:2019hou}. 
The following stellar evolution and supernova explosions can then produce NSs. For the NSs to remain embedded within the DS, their natal kick velocities (a recoil velocity from an asymmetric supernova) must be smaller than the escape velocity of the DS, $v_{\rm kick} < v_{\rm esc}
\simeq \left(2GM_{\rm DS}/R_{\rm DS}\right)^{1/2}$. Once embedded in the DS, the NSs lose orbital energy through DF and gradually 
move toward the central region, where they can form a bound NS binary. 

\item The second channel involves the dynamical capture of an NS by a pre-existing DS-NS system. When a second NS undergoes a close encounter with the DS--NS system, GW emission and DF can dissipate sufficient orbital energy to satisfy $\Delta E > \frac{1}{2}\mu v_{\infty}^2$, thereby producing a gravitationally bound configuration \cite{East:2012ww, Annulli:2020ilw, Boudon:2023vzl}. Subsequent dissipation drives the captured NS inward, while DF causes it to 
move toward the central NS. Eventually, the two NSs form a binary embedded within the DS.
\end{itemize}

\subsection{Physical assumptions and setup}\label{subsec:physical_assumptions}
In this section, we first discuss the physical assumptions underlying our future analysis, and then investigate the regime in which they are expected to hold. 

We consider an admixed configuration consisting of a central NS embedded within an extended, self-gravitating DS in hydrostatic equilibrium, as discussed in Sec.~\ref{Sect.Density_Profile}. 
In principle, the spacetime metric is determined by the total energy-momentum tensor of both matter components. Throughout this work, however, we will consider configurations that satisfy
\begin{equation*}
    \rho_{\rm DS,c}\ll\rho_{\rm NS,c} ~ ,\hspace{0.5cm}P_{\rm DS,c}\ll P_{\rm NS,c} ~ ,
\end{equation*}
where the subscript ``$\mathrm{c}$'' denotes central or core quantities. Under this assumption, the NS completely dominates the spacetime curvature near the origin. Consequently, the metric functions are well approximated by those of an isolated NS, that is
\begin{equation*}
    f(r)\simeq f_{\rm NS}(r) ~ ,\hspace{0.5cm}h(r)\simeq h_{\rm NS}(r) ~ ,
\end{equation*}
and the NS structure is effectively unchanged from the 
configuration of NS without DM. The DM component therefore acts as a medium surrounding the NS, allowing the NS spacetime to be treated as a fixed background in which the companion inspirals. The orbital dynamics are similarly only weakly affected by the presence of the DS. For a smooth DS density profile, the enclosed DM and NS mass near the center satisfy
\begin{equation*}
    M_{\rm DS}(r)\approx\frac{4\pi}{3}r^3\rho_{\rm DS,c}\hspace{0.5cm}M_{\rm NS}(r)\approx\frac{4\pi}{3}r^3\rho_{\rm NS,c},
\end{equation*}
so that close to the center we have $M_{\rm DS}(r)/M_{\rm NS}(r)\approx\rho_{\rm DS,c}/\rho_{\rm NS,c}\ll1$. Therefore, for the configurations we consider here $M_{\rm DS}(r)\ll M_{\rm NS}+M_*$, throughout the inspiral region inside the DS, where $M_*$ denotes the companion mass. As a result, the gravitational potential is dominated by the NS and the DM effects are negligible. Therefore, the orbital frequency 
can be safely described by the Kepler formula
\begin{equation}
    f_{\rm orb}=\frac{1}{2\pi}\sqrt{\frac{G(M_{\rm NS}+M_*)}{r^3}},
\end{equation}
with the influence of the DS entering only through the dissipative force associated with DF.

Meanwhile, as the inspiral proceeds, the companion and the central NS gradually approach one another, causing the position of the NS to change within the DS. The surrounding DM distribution must then readjust to the changing gravitational potential of the NS in order to remain approximately centered on it. The characteristic timescale for this readjustment can be estimated from the sound-crossing time, $t_s \sim L/c_s$, where $L$ is the characteristic length scale over which the DM density must redistribute. On the other hand, the timescale associated with the motion of the NS across the same region is $t_{\rm mot} \sim L/v_{\rm NS}$, where $v_{\rm NS}$ is the orbital velocity of the central NS around the binary center of mass. The condition $t_s < t_{\rm mot}$ ensures that pressure perturbations can propagate through the relevant region faster than the NS moves, allowing the DM distribution to respond to the motion of the NS. This condition can be equivalently written as $\mathcal{M}_{\rm NS}\equiv v_{\rm NS}/c_s<1$. For the DS configurations considered here, the central sound speed is $\sim0.1c$, while $v_{\rm NS}<c_s$ over the frequency range relevant to our analysis. Thus, the DS can continuously readjust and remain approximately centered on the NS throughout the inspiral.

\begin{figure}[htbp]
\centering 
\includegraphics[width=0.48\textwidth]{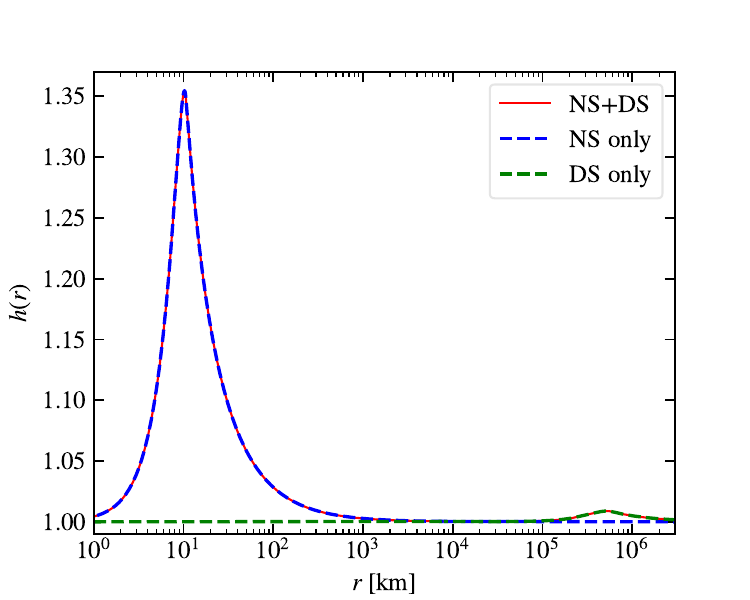}
\includegraphics[width=0.48\textwidth]{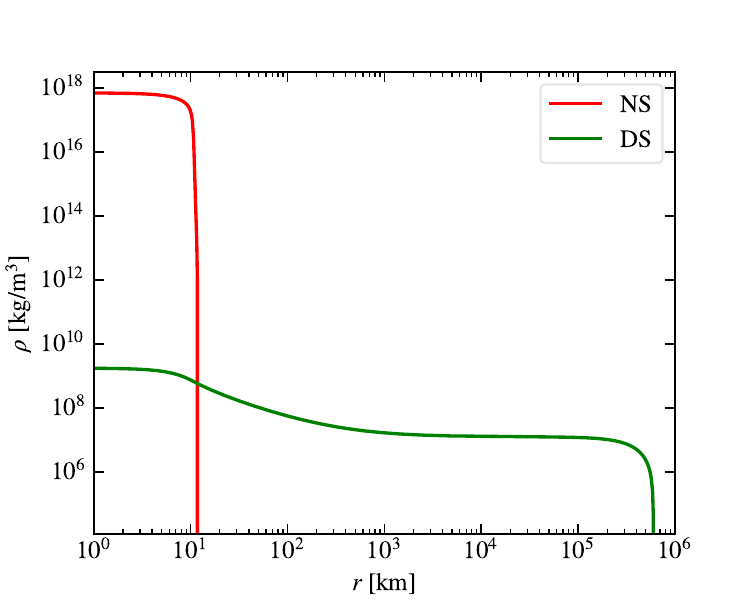}
\caption{\textit{Left panel:} Metric component $h(r)$ for a representative NS-DS configuration. The solid red curve shows the full solution, the dashed blue curve considers the NS component alone, and the dashed green curve considers the DS component alone. \textit{Right panel:} Density profiles corresponding to the solutions shown in the left panel. The solid red curve shows the NS density, while the green curve shows the DS density.} 
\label{fig:Metric}
\end{figure}

To illustrate the assumptions discussed above, we show a representative NS-DS configuration in Fig.~\ref{fig:Metric}. The left panel displays the metric function $h(r)$ for the full solution
, for the NS component alone
, and for the DS component alone
, denoted by solid red, dashed blue and dashed green, respectively. From the plot, we can see that in the inner region, the full solution is nearly indistinguishable from the NS-only curve, confirming that the spacetime is dominated by the NS at small radii. On the contrary, at larger radii, the extended DM envelope causes the full solution to deviate from the NS-only result. In comparison, the DS-only curve remains subdominant in the inner region, reflecting the small enclosed DM mass there. The right panel shows the NS (red) and DS (green) density profiles for the coupled NS-DS system shown in the left panel. The NS density is compact and falls off rapidly outside the NS radius, while the DS density is much smaller near the center but extends to much larger radii, forming a dilute envelope around the NS. This separation of scales justifies treating the NS as the dominant gravitational source and the DM as a background medium through which the companion moves.

\subsection{Density profile} 
\label{Sect.Density_Profile}
In this section, we establish the mapping between the EoS and the equilibrium density profile of the NS–DS system. This mapping provides the theoretical basis for reconstructing the DM EoS from the density-dependent DF signal discussed in later sections.

Considering the static and spherically symmetric space-time with the metric,
\begin{equation}\label{eq:line_element}
    ds^2=-f(r)c^2dt^2+h(r)dr^2+r^2(d\theta^2+\sin^2\theta d\phi^2) ~ , 
\end{equation}
and energy-momentum tensor
\begin{equation}\label{TMUNU}
    T^{\mu\nu}=P g^{\mu\nu}+\left(\frac{P}{c^2}+\rho\right)u^\mu u^\nu~,
\end{equation}
where $P$ and $\rho$ are the total pressure and energy density, respectively.

We denote the energy density and pressure of matter component $i$ as $\rho_i$ and $P_i$, respectively. For the NS-DS system, we then have the total energy density and pressure
\begin{align}
    \rho(r)&=\rho_{\rm NS}(r)+\rho_{\rm DS}(r),\cr
    P(r)&=P_{\rm NS}(r)+P_{\rm DS}(r).
\end{align}
From the field equations, we get 
\begin{align}\label{eq:structure_general}
    h^{-1}&=1-\frac{2 G m(r)}{r c^2},\cr
    \frac{dm}{dr}&=4\pi r^2 \rho ,\cr
    \frac{1}{2f}\frac{df}{dr}&=\frac{G\left[m(r)+4\pi r^3 P/c^2\right]}{c^2 r[r-2G m(r)/c^2]},\cr
    \frac{dP_i}{dr}&=-[P_i(r)+\rho_i(r)c^2]\frac{1}{2f}\frac{df}{dr},
\end{align}
where $m(r)$ denotes the total enclosed mass at radial coordinate $r$. The last equation is the Tolman-Oppenheimer-Volkoff (TOV) equation that determines the pressure distribution of each fluid. To solve for the structure, we only need to consider the system of ODEs for $m,P_{\rm NS},P_{\rm DS}$. The radius and mass of each sector are obtained with the boundary conditions $P_i(R_i)=0$, $M=m(\max\{R_{\rm NS},R_{\rm DS}\})$. Regarding the boundary condition at the center, we have the small radius expansions:
\begin{align}
    m &= \frac{4\pi}{3}\left(\rho_{\rm NS,c}+\rho_{\rm DS,c}\right)r^3 ~ , \\
    P_{\rm NS} &=P_{\rm NS,c}-\frac{2\pi G}{3c^4}\left[\left(\rho_{\rm NS,c}+\rho_{\rm DS,c}\right)c^2+3\left(P_{\rm NS,c}+P_{\rm DS,c}\right)\right]\left(\rho_{\rm NS,c}c^2+P_{\rm NS,c}\right)r^2 ~ ,\\
    P_{\rm DS} &=P_{\rm DS,c}-\frac{2\pi G}{3c^4}\left[\left(\rho_{\rm DS,c}+\rho_{\rm NS,c}\right)c^2+3\left(P_{\rm DS,c}+P_{\rm NS,c}\right)\right]\left(\rho_{\rm DS,c}c^2+P_{\rm DS,c}\right)r^2 ~ . 
\end{align}
To obtain the density profile by solving the equations above, we still need to know the EoS of both NS and the DS. We introduce these two EoSs in the following subsections.

\subsubsection{Neutron matter equation of state}
We consider the following parametrization for the NS EoS~\cite{Haensel:2004nu,Potekhin:2013qqa}:
\begin{equation}
\begin{aligned}
\zeta &= \frac{a_1+a_2\xi+a_3\xi^3}{1+a_4\xi}F[a_5(\xi-a_6)] + (a_7+a_8\xi)F[a_9(a_{10}-\xi)] \\
&\quad + (a_{11}+a_{12}\xi)F[a_{13}(a_{14}-\xi)] + (a_{15}+a_{16}\xi)F[a_{17}(a_{18}-\xi)] ~ , \label{eq:NS_EOS}\\[4pt]
\end{aligned}
\end{equation}
with
\begin{equation*}
\begin{alignedat}{5}
a_1 &= 6.2 ~ ,     &\quad
a_2 &= 6.121 ~ ,    &\quad
a_3 &= 0.005925 ~ , &\quad
a_4 &= 0.16326 ~ ,  &\quad
a_5 &= 6.48 ~ ,     \\
a_6 &= 11.4971 ~ ,  &
a_7 &= 19.105 ~ ,   &
a_8 &= 0.8938 ~ ,   &
a_9 &= 6.54 ~ ,     &
a_{10} &= 11.4950 ~ , \\
a_{11} &= -22.775 ~ , &
a_{12} &= 1.5707 ~ ,  &
a_{13} &= 4.3 ~ ,     &
a_{14} &= 14.08 ~ ,   &
a_{15} &= 27.8 ~ ,    \\
a_{16} &= -1.653 ~ , &
a_{17} &= 1.5 ~ ,    &
a_{18} &= 14.67 ~ .  &&
\end{alignedat}
\end{equation*}
where we define the function $F(x)\equiv \left(\mathrm{e}^x+1\right)^{-1}$, as well as the variables $\zeta\equiv\log(P_{\rm NS}/\mathrm{dyn}~\mathrm{cm}^{-2})$ and $\xi\equiv\log(\rho_{\rm NS}/\mathrm{g}~\mathrm{cm}^{-3})$. The coefficient choices correspond to the SLy EoS~\cite{Chabanat:1997qh,Douchin:2001sv}.

We note that the choice of the NS EoS is not central to the analysis presented here, as our primary focus is the structure and dynamics of the matter surrounding the NS. We adopt the SLy EoS as a representative, well-established model for the NS interior. We have verified that the results of interest are not sensitive to the particular choice of a realistic NS EoS, with differences in the stellar structure having only a negligible impact on the properties of the surrounding DM. We therefore use the SLy EoS throughout for definiteness.

\subsubsection{Dark matter equations of state}
In the following, we introduce two representative DM EoSs, corresponding to self-interacting bosonic and fermionic matter.

\paragraph{Bosonic EoS: } As one of our benchmark DM models, we consider a complex scalar field with a repulsive quartic self-interaction,
\begin{equation}
\mathcal{L}
=-\frac{1}{2}g^{\mu\nu}\nabla_\mu\phi^*\nabla_\nu\phi
-\frac{1}{2}m_\chi^2|\phi|^2
-\frac{\lambda}{4}|\phi|^4 ~ , 
\end{equation}
where $m_\chi$ is the boson mass and $\lambda>0$ is the self-coupling. In the strong self-interaction regime defined by~\cite{Colpi:1986ye}
\begin{equation}
\Lambda\equiv\frac{\lambda}{4\pi}\frac{M_{\rm Pl}^2}{m_\chi^2}\gg1 ~ ,
\end{equation}
the Einstein-Klein-Gordon system becomes equivalent to a perfect fluid obeying an algebraic EoS. In this limit, the scalar-field gradient energy is subdominant throughout most of the star, allowing the bosonic condensate to be described hydrodynamically~\cite{Colpi:1986ye}. The corresponding EoS can be written as
\begin{equation}
P_{\rm DS}(\rho_{\rm DS})=\frac{\rho_0 c^2}{36\pi}\left[\sqrt{1+12\pi\left(\frac{\rho_{\rm DS}}{\rho_0}\right)}-1\right]^2,
\label{eq:EoSBS}
\end{equation}
where we define $\rho_0\equiv 2.92\times10^{18}\,\mathrm{g}~\mathrm{cm}^{-3}\,\lambda^{-1}\left(m_\chi/1~\mathrm{GeV}\right)^4$. The resulting boson star solutions exhibit a scaling symmetry, depending only on the combination $\lambda^{-1} m_\chi^{4}$ rather than on $\lambda$ and $m_\chi$ separately. With this in mind, we shall henceforth fix the self-coupling to $\lambda=0.01$ and only vary the boson mass to consider different configurations.

\paragraph{Fermionic EoS: } As a second benchmark model, we consider fermionic DM. In this case, the pressure supporting the star against gravitational collapse originates from Fermi degeneracy pressure. We further allow for self-interactions mediated by a massive boson of mass $m_\phi$ with coupling strength $\alpha=g^2/(4\pi)$. Following the mean-field treatment of Ref.~\cite{Kouvaris:2015rea}, the EoS is conveniently written in parametric form in terms of the dimensionless Fermi momentum $x\equiv p_F/m_\chi$, where $m_\chi$ is now the fermion mass. The energy density and pressure are
\begin{subequations}\label{eq:EoSFS}
\begin{align} \rho_{\rm DS}(x)&=\frac{g_S c^3}{2\hbar^3}m_\chi^4 \left[ \xi(x) \pm \frac{2\alpha}{9\pi^3} \frac{g_S}{2} \left(\frac{m_\chi}{m_\phi}\right)^2 x^6 \right],\\
P_{\rm DS}(x)&=\frac{g_S c^5}{2\hbar^3}m_\chi^4 \left[ \psi(x) \pm \frac{2\alpha}{9\pi^3} \frac{g_S}{2} \left(\frac{m_\chi}{m_\phi}\right)^2 x^6 \right],
\end{align}
\end{subequations} 
where $g_S=2S+1$ is the spin multiplicity. The upper (lower) sign corresponds to repulsive (attractive) self-interactions. In this paper, we will consider only repulsive self-interactions. The functions 
\begin{align} 
\xi(x)&=\frac{1}{8\pi^2}\left[x\sqrt{1+x^2}(1+2x^2)-\ln\!\left(x+\sqrt{1+x^2}\right)\right],\\ \psi(x)&=\frac{1}{8\pi^2}\left[x\sqrt{1+x^2}\left(\frac{2x^2}{3}-1\right)+\ln\!\left(x+\sqrt{1+x^2}\right)\right]  ~ ,
\end{align} 
are the standard zero-temperature Fermi-Dirac integrals. Unlike the bosonic EoS, the fermionic EoS cannot be expressed analytically in the form $P(\rho)$ and must instead be specified parametrically through $x$. We also note that the stellar structure depends only on the two combinations $m_\chi$ and $\alpha/m_\phi^2$, rather than on the three microphysical parameters independently. As in the boson-star case, we will henceforth fix $\alpha/m_\phi^2=10^{-3}~\mathrm{MeV}^{-2}$.

\section{Dynamical Friction and $D$--Function}
\label{Sect.DynamicalFriction}

In this section, we briefly introduce how GW observations for binary system evolution can be used to extract information about a self-interacting cloud, following Ref.~\cite{Ding:2025hqf}.

\subsection{$D$--function}
We consider a compact object embedded in a dense dark environment, such as a self-interacting boson cloud, with a binary companion moving inside the cloud. In addition to GW emission, the orbital motion also experiences an extra energy loss due to DF~\cite{1943ApJ....97..255C}. This additional power loss modifies the evolution of the GW frequency and can be related to the local density of the cloud at the position of the compact object: 
    \begin{equation}\label{eq:P_DF}
        P_{\rm DF}=-4\pi \frac{G^2M_*^2}{v}\rho C_\Lambda ~ .
    \end{equation}
Here, $M_\ast$ denotes the mass of the binary companion, $v\equiv(GM_{\rm tot}/r)^{1/2}$ is the relative velocity between the companion and the dark environment, where $M_{\rm tot}$ is the total mass of the binary and $r$ is the binary separation. $\rho$ is the local density of the dark environment experienced by the companion, and $C_\Lambda$ is the Coulomb logarithm. The value of $C_\Lambda$ is determined by the properties of the cloud, as will be discussed in detail in Sec.~\ref{Sect.CLambda}. 

The GW frequency $f$ in the observer rest frame evolves due to both GW emission and the power loss induced by DF, which can be expressed as
\begin{equation}\label{Eq.dfdt}
    \frac{df}{dt} = \frac{96}{5}\frac{\left[G\mathcal M_c(1+z)\right]^{5/3}\pi^{8/3}f^{11/3}}{c^5}+\frac{3 f^{1/3}}{(\pi G)^{2/3}[\mathcal M_c(1+z)]^{5/3}}|P_{\rm DF}| ~ , 
\end{equation}
where the first term is the GW emission contribution, while the second one is the DF part. Here $\mathcal M_c\equiv(MM_*)^{3/5}/(M+M_*)^{1/5}$ is the chirp mass, and $z$ is the cosmic redshift, both of which can be determined by early phase GW observation. From Eq.~\eqref{Eq.dfdt}, we can see that the contribution of $P_{\rm DF}$ term, with the information of the environment encoded, will cause GW frequency dephasing.

To obtain information about the surrounding dark environment, we need to extract information from $P_{\rm DF}$, we then introduce another quantity $D$ that is fully constructed by GW observables, GW amplitude $h$, GW frequency $f$ and their time evolution  (see Appendix.~\ref{app:derivation_D} and Ref.~\cite{Ding:2025hqf} for detailed derivation)
\begin{equation}
    D\equiv\frac{dh}{dt} + 3 \frac{h}{f} \frac{df}{dt} - \frac{h}{df/dt}\frac{d^2f}{dt^2}=\frac{12G}{c^4 d_L(z)}|P_{\rm DF}|\left(\frac{11}{3}-\frac{d\ln\rho}{d\ln f}-\frac{ d\ln C_\Lambda}{d\ln f}\right) ~ .   
    \label{Eq.D}
\end{equation}
Here, $d_L(z) \equiv (1+z) \int_0^z c/H(z') dz'$ is the luminosity distance between the compact object and the observer, and $H(z) = H_0 (\Omega_m (1+z)^3 +\Omega_\gamma(1+z)^4 +\Omega_\Lambda)^{1/2}$ is the Hubble parameter as a function of redshift in $\Lambda$CDM model with cosmological parameters $(H_0,\, \Omega_m,\, \Omega_\gamma,\, \Omega_\Lambda)$ from Planck 2018 \cite{Planck:2018vyg}. 
If the primary is not surrounded by a dark environment and the orbital shrinkage of the binary companion is driven purely by GW emission without extra power loss, then we have $D=0$. Otherwise, the additional power loss due to DF from the dark environment will cause $D$ to vary with the GW frequency $f$.
Then, the density profile $\rho(r)$ of the DS can be reconstructed by combining the density-profile ansatz introduced in Sec.~\ref{Sect.Density_Profile} with the observational information from the $D$ function, which will be discussed in Sec.~\ref{Sect.Cloud_Properties}.

\subsection{Coulomb logarithm}
\label{Sect.CLambda}

The Coulomb logarithm $C_\Lambda$ in Eq.~\eqref{eq:P_DF} depends on the properties of the cloud, such as the cloud size and the Mach number 
\begin{equation}
\mathcal M \equiv v/c_s ~ , 
\label{Eq.MachNo}     
\end{equation} 
where $c_s$ is the sound speed determined by the equation of state (EoS) of the medium.

In \cite{Berezhiani:2023vlo}, the authors studied DF in a superfluid medium, which is a good approximation for a self-interacting cloud in a DS. In their work, the DF force is given by
\begin{equation}
    F_{\rm DF} = -\frac{4\pi G^2\mu^2\rho}{c_s^2} \mathcal F_{\rm DF} ~ , 
\end{equation} 
where $\mathcal F_{\rm DF}$ is the dimensionless force in the form of a sum over angular multipoles $(l,m)$ 
\begin{equation}
    \mathcal F_{\rm DF} \equiv c_s^2\sum_{l=1}^{l_{\rm max}}\sum_{m=-l}^{l-2}\gamma_{lm} \left\{{\rm Re} \left(S^m_{l,l-1}-S^{m+1}_{l,l-1}{}^*\right)\hat r + {\rm Im}\left(S_{l,l-1}^m-S_{l,l-1}^{m+1}{}^*\right) \hat\varphi \right\}~ , 
    \label{Eq.dimlessFDF}
\end{equation}
where
\begin{equation}
    \gamma_{lm} = (-1)^m \frac{(l-m)!}{(l-m-2)!}\left\{\Gamma\left(\frac{1-l-m}{2}\right)\Gamma\left(1+\frac{l-m}{2}\right)\Gamma\left(\frac{3-l+m}{2}\right)\Gamma\left(1+\frac{l+m}{2}\right)\right\}^{-1} ~ .   
\end{equation}
Here, $\Gamma$ denotes the Gamma function, and  $l_{\rm max}\sim1.2/(r_{\rm min}/R_p)$, where $R_p$ is the binary separation, and $r_{\rm min}$ is the minimum length scale down to which the point-particle and linear-fluid descriptions remain valid, usually quantified by the radius of the binary companion. Since only the azimuthal component of the DF force is relevant for the orbital evolution, we define $\mathcal F_\varphi \equiv \mathcal F_{\rm DF} \cdot \hat\varphi$ and $F_\varphi \equiv F_{\rm DF} \cdot \hat\varphi$. 

In the subsonic region with $\mathcal M<1$, the resummed version takes the exact form 
\begin{equation}
    F_\varphi = -\frac{4\pi G^2\mu^2\rho}{v^2} ({\rm arctanh} \, \mathcal M-\mathcal M) ~ .  
\end{equation}
On the other hand, in the supersonic 
sound regime, we have the momentum integral in Eq.~\eqref{Eq.dimlessFDF} simplified to \cite{2022ApJ...928...64D}
\begin{equation}
    S^m_{l,l-1} = \frac{i\pi}{2c_s^2} j_l(m\mathcal M)h_{l-1}^{(1)}(m\mathcal M) ~ ,   
    \label{Eq.Sml}
\end{equation}
where $j_l$ and $h_l^{(1)}$ are the spherical Bessel and Hankel functions, respectively, together with 
\begin{equation}
    S_{l,l-1}^0 = \frac{\pi}{2c_s^2(4l^2-1)} ~ . 
    \label{Eq.S0l}
\end{equation} 

\begin{figure}[htbp]
\centering
\includegraphics[width=10cm]{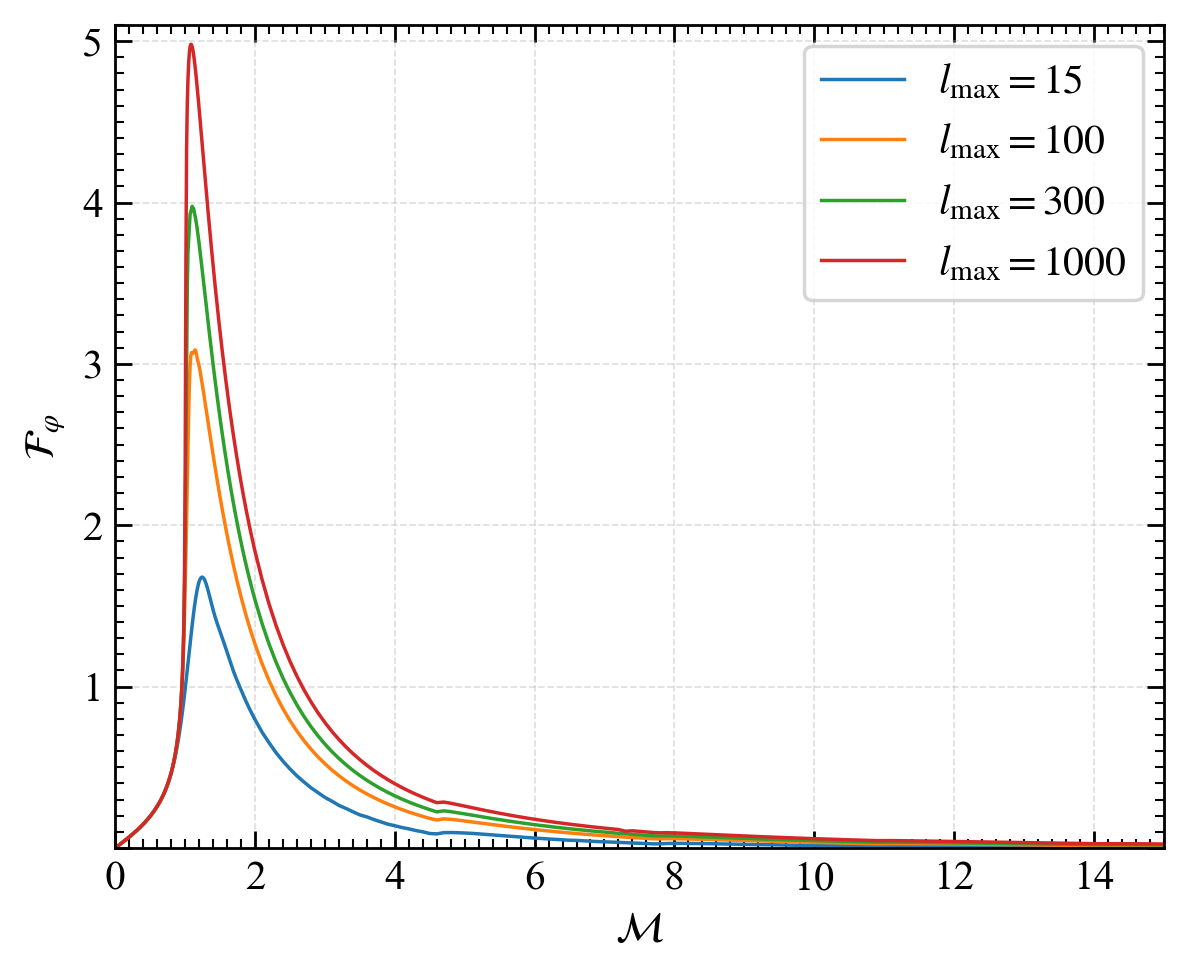}
\caption{The azimuthal component of the dimensionless DF force $\mathcal F_\varphi$ as a function of Mach number $\mathcal M$. The blue, orange, green and red line indicate $l_{\rm max}=15$, $100$, $300$ and $1000$ respectively, which indicates different binary separation. }
\label{Fig.lKl}
\end{figure}

Taking Eq.~\eqref{Eq.Sml} and Eq.~\eqref{Eq.S0l} into Eq.~\eqref{Eq.dimlessFDF}, we obtain the relation between Mach number $\mathcal{M}$ and the dimensionless DF $\mathcal F_{\rm DF}$. The relation between the azimuthal component of the dimensionless DF force $\mathcal F_\varphi$ and $\mathcal M$ is shown in Fig.~\ref{Fig.lKl}, and the Coulomb logarithm $C_\Lambda$ can be expressed as 
\begin{equation}
    C_\Lambda (\mathcal M) = \mathcal M^2\mathcal F_\varphi(\mathcal M) ~ .     
\end{equation}

Thus, we numerically evaluate the kernel function at relatively small $l$ and use its asymptotic behavior to extrapolate to large $l$. We then perform the summation to obtain $\hat{\mathcal F}_{\rm DF}\cdot \hat\varphi$. The result is shown in Fig.~\ref{Fig.lKl}.

\section{Equation of State Imprints on Dark Star Structure and GW Signals}\label{sec:examples_DS}

Having established the general numerical framework for determining the hydrostatic structure of a DS in Sec.~\ref{Sect.Density_Profile} and its corresponding GW observable $D(f)$ in Sec.~\ref{Sect.DynamicalFriction}, we now investigate how the DS EoS is imprinted on the resulting density profile and GW signal. Our goal in this section is to develop physical intuition for the forward mapping from the DM EoS to $D(f)$, which is the basis of the reconstruction procedure introduced in Sec.~\ref{Sect.Cloud_Properties}. 

\subsection{Analytical approximation}
In this section, we consider an approach that provides analytic insight into the relationship between the DM EoS, the density profile, and the resulting DF signal.

For the bosonic case, the low-density limit of Eq.~\eqref{eq:EoSBS} is a relation given by $P_{\rm DS}\propto \rho_{\rm DS}^2$. For a polytropic parametrization of the form $P_{\rm DS}\propto \rho_{\rm DS}^{1+1/n}$, this corresponds to a polytropic index $n=1$. For the fermionic case, the asymptotic low-density limit of Eq.~\eqref{eq:EoSFS} is instead the usual degenerate Fermi-gas relation, $P_{\rm DS}\propto\rho_{\rm DS}^{5/3}$. However, for sufficiently strong repulsive self-interactions, there exists an intermediate density regime in which the interaction contribution dominates the pressure while the energy density remains dominated by the free-fermion contribution. In this regime, the fermionic EoS also takes an approximately polytropic form, $P_{\rm DS}\propto \rho_{\rm DS}^2$. Therefore, for the two DM models considered in this work, we have
\begin{equation}
\begin{cases}
    P_{\rm DS}\approx\frac{c^2\pi}{\rho_0}\rho_{\rm DS}^2 ~ ,\hspace{2.45cm}\text{self-interacting bosons} ,\cr
    P_{\rm DS}\approx\frac{2\pi \hbar^3}{c}\frac{\alpha}{(m_{\chi}m_{\phi})^2}\rho_{\rm DS}^2 ~ ,\hspace{0.94cm}\text{self-interacting fermions}.
\end{cases}
\end{equation}
Here the bosonic expression is the true low-density limit, while the fermionic expression applies only in the self-interaction-dominated intermediate regime (at even lower densities we recover $P_{\rm DS}\propto\rho_{\rm DS}^{5/3}$). For this polytropic EoS, and under the assumptions that: i) the NS structure is not substantially modified by the DM and ii) the DM density is low, the coupled two-fluid TOV equations can be approximated as a constant-density NS and a DS that follows the non-relativistic hydrostatic equilibrium equation. This setup admits an exact analytical solution for the density profile, allowing us to derive closed-form expressions for the DS radius, mass, and the associated $D$-function. This analytic solution not only serves as a cross-check for our numerical methodology, but also serves as a window into the qualitative behavior of the signal which reveals how the $D$-function depends on the frequency and on stellar parameters.

\paragraph{Density profile:} The simplified system of differential equations we solve is:
\begin{align}\label{eq:approx_structure}
    \frac{dm}{dr}&=4\pi r^2 (\rho_{\rm NS}+\rho_{\rm DS}) ~ ,\cr
    \frac{dP_{\rm NS}}{dr}&=-(\rho_{\rm NS}+P_{\rm NS})\frac{4\pi r^3 P_{\rm NS}+m_{\rm NS}}{r(r-2m_{\rm NS})} ~ ,\cr
    \frac{dP_{\rm DS}}{dr}&=-\frac{Gm\rho_{\rm DS}}{r^2} ~ ,
\end{align}
where for the NS pressure equation we have approximated the enclosed mass $m\approx m_{\rm NS}$ and $P\approx P_{\rm NS}$. Under this approximation, the structure of the NS is then explicitly decoupled from the DM and can be solved independently, which is in line with our requirement that the NS is not substantially modified by the addition of DM. We additionally assume that this matter is modeled by a constant density $\rho_{\rm NS}$ for ease of calculation, which is given by $\rho_{\rm NS}=3M_{\rm NS}/(4\pi R_{\rm NS}^3)$ with $M_{\rm NS},\,R_{\rm NS}$ the total NS mass and radius, respectively. On the other hand, the DS follows the non-relativistic hydrostatic equilibrium equation, where the NS contribution enters through the enclosed mass term $m=m_{\rm NS}+m_{\rm DS}$.

Taking the polytropic form $P_{\rm DS}=K\rho_{\rm DS}^2$ for the DM, we obtain the structure equation for the density $d\rho_{\rm DS}/dr=-Gm/(2Kr^2)$. Additionally, expressing the DS density as $\rho_{\rm DS}=(dm_{\rm DS}/dr)/(4\pi r^2)$ and differentiating yields the following differential equation for the DS mass
\begin{equation}
    \frac{d^2m_{\rm DS}}{dr^2}-\frac{2}{r}\frac{dm_{\rm DS}}{dr}+\frac{m_{\rm DS}}{\alpha^2}=-\frac{m_{\rm NS}}{\alpha^2} ~ ,
\end{equation}
where we have written $m=m_{\rm NS}+m_{\rm DS}$ and defined $\alpha^2\equiv K/(2\pi G)$. Given the linearity of the equation, we express its solution as the sum of the homogeneous solution and a particular solution. The solution of the homogeneous equation is written in terms of the spherical Bessel functions of first order:
\begin{equation}
    m_{\rm DS}^{\rm hom}=A\alpha[\sin(r/\alpha)-(r/\alpha)\cos(r/\alpha)]-B\alpha[\cos(r/\alpha)+(r/\alpha)\sin(r/\alpha)] ~ ,
\end{equation}
with $A,B=\rm const$. The particular solution depends on the behavior of the source term $-m_{\rm NS}/\alpha^2$. For $r>R_{\rm NS}$ it is simply given by $-M_{\rm NS}/\alpha^2=\rm const$, while for $r<R_{\rm NS}$ it takes the form $-4\pi\rho_{\rm NS}r^3/(3\alpha^2)$. This gives the particular solutions:
\begin{equation}
m_{\rm DS}^{\rm part}=
    \begin{cases}
        -\frac{4\pi}{3}\rho_{\rm NS}r^3 ~ ,\hspace{0.5cm} \text{if }r<R_{\rm NS} ~ ,\cr
        -M_{\rm NS} ~ ,\hspace{1.1cm} \text{if }r\geq R_{\rm NS} ~ .
    \end{cases}
\end{equation}
Therefore, we can write the solution as the solution for $r<R_{\rm NS}$ and also for $r\geq R_{\rm NS}$.

The integration constants are determined by combining the regularity conditions at the center with the continuity requirements at the NS surface. At the origin, we demand $m_{\rm DS}(0)=0$ and that the small-$r$ behavior reproduces a finite central DM density, $m_{\rm DS}(r)\simeq \frac{4\pi}{3}\rho_{\rm DS,c}\,r^3$ as $r\to0$. This fixes the amplitude of the inner homogeneous solution in terms of the central density $\rho_{\rm DS,c}$. Across the NS boundary at $r=R_{\rm NS}$ both the enclosed DS mass and its first derivative must be continuous, that is $m_{\rm DS}^{\rm inner}(R_{\rm NS})=m_{\rm DS}^{\rm outer}(R_{\rm NS})$ and ${m_{\rm DS}^{\rm inner}}'(R_{\rm NS})={m_{\rm DS}^{\rm outer}}'(R_{\rm NS})$. These matching conditions then determine the two constants $A_{\rm out}$ and $B_{\rm out}$ of the outer solution, leaving $\rho_{\rm DS,c}$ as the sole remaining free parameter. We then obtain:
\begin{equation}
    m_{\rm DS}(\xi)=
    \begin{cases}
        4\pi\alpha^3(\rho_{\rm NS}+\rho_{\rm DS,c})[\sin\xi-\xi\cos\xi]-\frac{4\pi}{3}\rho_{\rm NS}\alpha^3\xi^3 ~ ,\hspace{1.8cm} \text{if }\xi<\xi_{\rm NS} ~ ,\cr
        4\pi\alpha^3[A_{\rm out}(\sin\xi-\xi\cos\xi)+B_{\rm out}(\cos\xi+\xi\sin\xi)]-M_{\rm NS} ~ ,\hspace{0.25cm} \text{if }\xi\geq \xi_{\rm NS} ~ ,
    \end{cases}
\end{equation}
where we have defined the variable $\xi\equiv r/\alpha$ ($\xi_{\rm NS}\equiv R_{\rm NS}/\alpha$) and the integration constants are
\begin{align}
    A_{\rm out}&=\rho_{\rm DS,c}+\rho_{\rm NS}(1-\cos\xi_{\rm NS}-\xi_{\rm NS}\sin\xi_{\rm NS}) ~ ,\cr
    B_{\rm out}&=\rho_{\rm NS}[\sin\xi_{\rm NS}-\xi_{\rm NS}\cos\xi_{\rm NS}] ~ .
\end{align}
From here, we calculate the DS density profile by taking $\rho_{\rm DS}=(dm_{\rm DS}/dr)/(4\pi r^2)$:
\begin{equation}\label{Eq.AnaRho}
    \rho_{\rm DS}(\xi)=
    \begin{cases}
        (\rho_{\rm NS}+\rho_{\rm DS,c})\frac{\sin\xi}{\xi}-\rho_{\rm NS} ~ ,\hspace{1.1cm} \text{if }\xi<\xi_{\rm NS} ~ ,\cr
        A_{\rm out}\frac{\sin\xi}{\xi}+B_{\rm out}\frac{\cos\xi}{\xi} ~ ,\hspace{1.76cm} \text{if }\xi\geq \xi_{\rm NS} ~ .
    \end{cases}
\end{equation}
Both the mass and density profiles are only valid until the DS radius $R_{\rm DS}=\alpha\xi_{\rm DS}$, which is implicitly defined as $\rho_{\rm DS}(\xi_{\rm DS})=0$. From Eq.~\eqref{Eq.AnaRho}, we see it can be written as 
\begin{equation}\label{eq:DM_radius_approx}
    R_{\rm DS}=\alpha\xi_{\rm DS}=\alpha\arctan\left(-\frac{B_{\rm out}}{A_{\rm out}}\right) ~ ,
\end{equation}
under the \textit{a-priori} assumption that $\xi_{\rm DS}>\xi_{\rm NS}$, which is our case of interest. We then understand that for $\xi>\xi_{\rm DS}$ the DM density vanishes and that the total DS mass is given by $m_{\rm DS}(\xi_{\rm DS})$.

\begin{figure}[htbp]
\centering 
\includegraphics[width=0.48\textwidth]{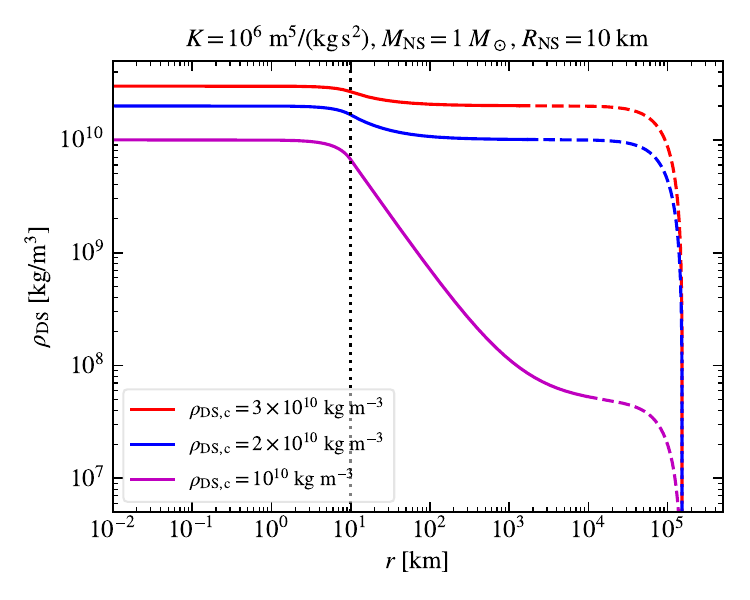} 
\includegraphics[width=0.48\textwidth]{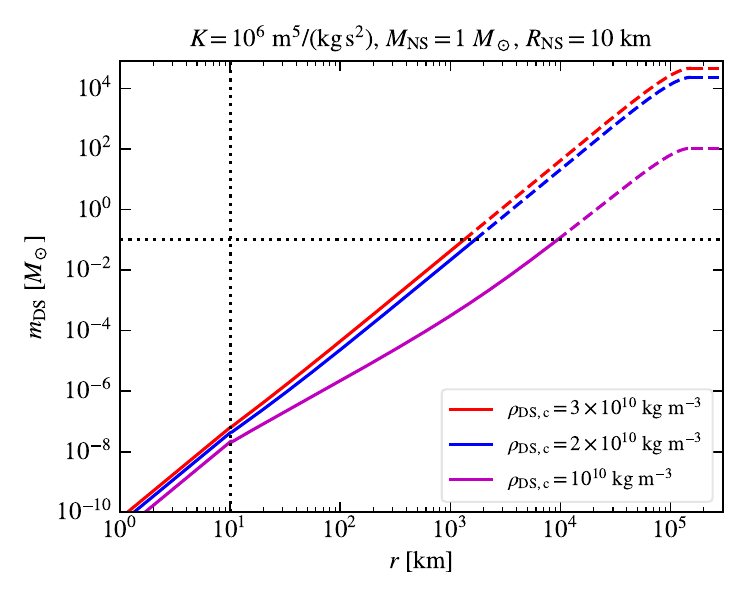} 
\caption{\textit{Left panel:} DS density profiles for $K=10^6~\mathrm{m}^5/(\mathrm{kg}\,\mathrm{s}^2)$ and three different DS core densities. \textit{Right panel:} Enclosed DS mass for the profiles in the left panel. In all panels the NS mass is $1\,M_\odot$ and the radius is $10$ km; it is represented by the vertical dashed line. The dashed colored curves represent the segments where the DS mass is larger than $0.1\,M_\odot$ and the approximate Kepler formula is no longer valid.} 
\label{fig:ApproxProfiles}
\end{figure}

In Fig.~\ref{fig:ApproxProfiles}, we present example DS density profiles in the left panel and the related enclosed DS mass in the right panel. In both panels, the solid curve segments represent the radii where the enclosed DS mass is smaller than $0.1\,M_\odot$, corresponding to the radii where the physical assumptions from Sec.~\ref{subsec:physical_assumptions} are satisfied. In contrast, the dashed segments show the radii where the approximate Kepler formula is invalid and outside of the scope of our analysis.

We note from the density profiles that, in this approximation, the DS radius is almost independent of the core DS density. This is consistent with Eq.~\eqref{eq:DM_radius_approx}, as the ratio $B_{\rm out}/A_{\rm out}$ depends weakly on $\rho_{\rm DS,c}$ in the limit $\rho_{\rm DS,c}\ll\rho_{\rm NS}$. However, we see by comparing the magenta curve with the red and blue lines that raising the value of $\rho_{\rm DS,c}$ yields flatter density profiles. This leads to a substantial increase of the DS mass of over two orders of magnitude. For the smallest core density we consider, we can still observe the formation of a ``plateau'' region at large radii, however, for even smaller $\rho_{\rm DS,c}$, the density profile will decay completely without forming a constant-density region. For small enough core density the DS radius will fall below the NS radius, and we will not be able to extract any information on the DM with our considerations. Finally, we note that the DS core density cannot be chosen to be arbitrarily large, since the speed of sound must be smaller than the speed of light in order to have physically valid solutions. For example, for $\rho_{\rm DS,c}=3\times10^{10}~\mathrm{kg}~\mathrm{m}^{-3}$, the speed of sound in the DM medium at the core would already be $c_s=\sqrt{2K\rho_{\rm DS,c}}\approx0.82c$.

\paragraph{$D$-function:} According to Eq.~\eqref{Eq.D}, the $D$-function depends on the Coulomb logarithm and its derivative. In the supersonic region, its expression is complicated for approximate analysis. In the subsonic region, however, it has the simple form $C_\Lambda=1/2\ln((1+\mathcal{M})/(1-\mathcal{M}))-\mathcal{M}$ \cite{Ostriker:1998fa}. We therefore work in this regime ($\mathcal{M}\ll1$). We are interested in the expression for $d\ln C_\Lambda/d\ln f$, so we first note that
\begin{equation}\label{eq:dCLambda_dM}
    \frac{d\ln C_\Lambda}{d\ln \mathcal{M}}=\frac{\mathcal{M}^3}{C_\Lambda(1-\mathcal{M}^2)} ~ .
\end{equation}
Furthermore, by the Keplerian relation ($r\propto f^{-2/3}$) and the definition of $\xi$, we have $d\xi/df=-2\xi/(3f)$. On the other hand, assuming we have circular motion gives an orbital velocity $v\propto f^{1/3}$, so the Mach number is proportional to $\mathcal{M}=v/c_s\propto f^{1/3}/c_s$, where $c_s^2=\partial P_{\rm DS}/\partial \rho_{\rm DS}$ is the sound speed of the DM. For a barotropic DM fluid ($P_{\rm DS}=P_{\rm DS}(\rho_{\rm DS})$), we have the logarithmic derivative
\begin{align}\label{eq:log_diffs}
    \frac{d\ln\mathcal{M}}{d\ln f}&=\frac{1}{3}-\frac{1}{2}\frac{d\ln c_s^2}{d\ln f}=\frac{1}{3}-\frac{1}{2}\left(\frac{d\ln c_s^2}{d\ln \rho_{\rm DS}}\right)\frac{d\ln \rho_{\rm DS}}{d\ln f} ~ .
\end{align}
Using the chain rule and eqs.~\eqref{eq:dCLambda_dM} and~\eqref{eq:log_diffs}, we can then write the derivative
\begin{equation}\label{eq:dlnC_L_dlnf_subsonic1}
    \frac{d\ln C_\Lambda}{d\ln f}=\frac{\mathcal{M}^3}{3C_\Lambda(1-\mathcal{M}^2)}\left[1-\frac{3}{2}\left(\frac{d\ln c_s^2}{d\ln \rho_{\rm DS}}\right)\frac{d\ln \rho_{\rm DS}}{d\ln f}\right].
\end{equation}
This is a general result for the subsonic regime, namely, it accounts for any barotropic form of the EoS and induced density profile. In the low-Mach limit $\mathcal{M}\ll1$, we can further approximate $C_\Lambda=\mathcal{M}^3/3+\mathcal{O}(\mathcal{M}^5)$, which reduces the expression above into
\begin{equation}\label{eq:dlnC_L_dlnf_subsonic2}
    \frac{d\ln C_\Lambda}{d\ln f}\approx1-\frac{3}{2}\left(\frac{d\ln c_s^2}{d\ln \rho_{\rm DS}}\right)\frac{d\ln \rho_{\rm DS}}{d\ln f} ~ .
\end{equation}
Furthermore, from the structure equations~\eqref{eq:approx_structure}, we can write
\begin{equation*}
    \frac{d\ln\rho_{\rm DS}}{d\ln f}=\frac{2Gm}{3\,r\,c_s^2}\approx\frac{2GM_{\rm NS}}{3\,r\,c_s^2} ~ ,
\end{equation*}
where in the last expression we have approximated the total mass as the NS mass only. With this, we can then write the approximate $D$-function
\begin{align}\label{eq:D_function_barotropic}
    D(f)&\approx\frac{8G}{c^4 d_L(z)}|P_{\rm DF}|\left[4-\left(1-\frac{3}{2}\left(\frac{d\ln c_s^2}{d\ln \rho_{\rm DS}}\right)\right)\frac{GM_{\rm NS}}{r(f)\,c_s^2}\right]\cr
    &=\frac{8G}{c^4 d_L(z)}|P_{\rm DF}|\left[4-\left(\frac{2n-3}{2(n+1)}\right)\frac{GM_{\rm NS}}{K r(f)\rho_{\rm DS}^{1/n}(r(f))}\right]~ ,   
\end{align}
where in the last equality we have particularized to a general polytropic DM EoS $P_{\rm DS}=K\rho_{\rm DS}^{1+1/n}$. These last two equations for the $D$-function are general for any EoS and any polytrope, respectively, under the assumption $\mathcal{M} \ll 1$. Now, we particularize to the $n=1$ polytrope, which we have solved analytically. For the $D$-function, we are mainly interested in the outer solution of the DM profile, since this is the region where the inspiral phase happens. Therefore, we will denote the DS density as only the outer part
\begin{equation*}
    \rho_{\rm DS}(\xi)=
    \begin{cases}
        A_{\rm out}\frac{\sin\xi}{\xi}+B_{\rm out}\frac{\cos\xi}{\xi},\hspace{0.5cm}\text{if }\xi<\xi_{\rm DS} ~ ,\cr
        0\hspace{3.75cm}\text{if }\xi\geq\xi_{\rm DS} ~ .
    \end{cases}
\end{equation*}
With this density distribution and related EoS, we then write the approximate $D$-function as
\begin{equation}\label{eq:D_func_approx}
    D(f)\approx\frac{4\pi G(G^2M_*)^2}{3c^4 d_L(z)K^3} \left(\frac{v(f)}{\rho_{\rm DS}(\xi(f))}\right)^2\left[4+\frac{GM_{\rm NS}}{4K \alpha\xi(f)\rho_{\rm DS}(\xi(f))}\right] ~ .
\end{equation}
As can be observed from Fig.~\ref{fig:ApproxProfiles}, the density profile is expected to have segments where it is close to constant, in particular at large $r$. In that case, the second term in the previous equation becomes subdominant and the $D$-function has a scaling with frequency of the form $D\propto v^2\propto f^{2/3}$.
\begin{figure}[htbp]
\centering 
\includegraphics[width=10cm]{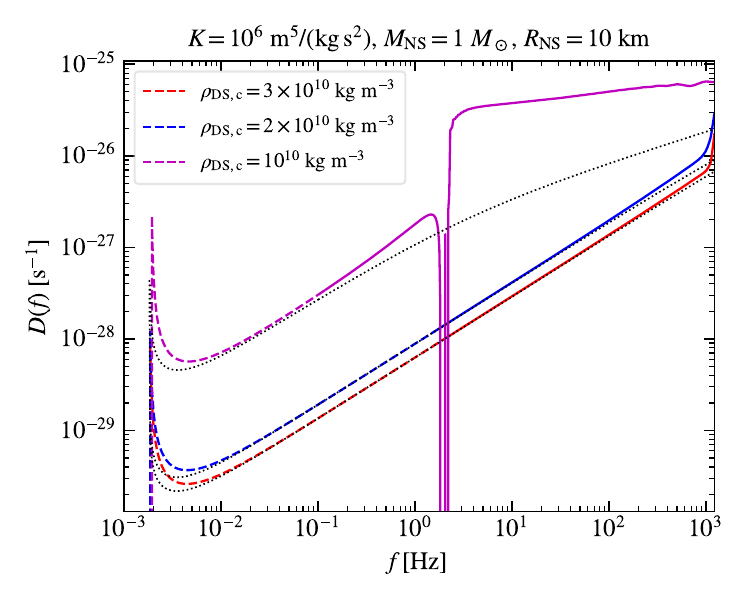} 
\caption{$D$-function as a function of $f$ for the cases shown in Fig.~\ref{fig:ApproxProfiles}. The mass of the perturber was chosen to be $M_*=1 \,M_\odot$. The colored lines correspond to the full calculation using Eq.~\eqref{Eq.D}, while the dotted black curves are obtained with the approximate formula from Eq.~\eqref{eq:D_func_approx}.  The dashed colored curves represent the segments where the DS mass is larger than $0.1 \,M_\odot$. The redshift in luminosity distance is set as $z=0.5$ as a benchmark value in calculation.} 
\label{fig:ApproxD_function}
\end{figure}

In Fig.~\ref{fig:ApproxD_function} we present the $D$-function calculated for the density profiles of Fig.~\ref{fig:ApproxProfiles}. The colored curves are obtained by using the full expression given by Eq.~\eqref{Eq.D}, while the dotted black curves use the approximation of Eq.~\eqref{eq:D_func_approx}. Regarding the red and blue curves, we see that Eq.~\eqref{eq:D_func_approx} is a very good approximation for most frequencies. The reason for this is that the amplitude of the density for these two solutions remains large for most radii, which means $c_s\sim c$ and so the Mach number $\mathcal{M}\ll1$ for most of the inspiral, which is the precise regime where Eqs.~\eqref{eq:dlnC_L_dlnf_subsonic1} and~\eqref{eq:dlnC_L_dlnf_subsonic2} are valid. Additionally, we note that the scaling $D\propto f^{2/3}$ is satisfied throughout most frequencies. This is due to the fact that the density profiles exhibit a clear plateau region that yields $d\ln\rho_{\rm DS}/d\ln f\approx0$, which makes the second term in the bracket of Eq.~\eqref{eq:D_func_approx} subdominant.

On the other hand, the magenta curve only fits the behavior of Eq.~\eqref{eq:D_func_approx} at small frequencies, referring to large radii. However, the approximation breaks down for $f\gtrsim1~\rm Hz$ ($r\lesssim 2200~\rm km$). The reason is twofold: i) there is a rise in the density, such that $d\ln\rho_{\rm DS}/d\ln f\neq0$ and most importantly, ii) the amplitude of the density is small enough such that the speed of sound in the medium is rapidly overtaken by the circular velocity, hence the perturber's orbit enters the supersonic regime $\mathcal{M}>1$. Both effects lead to the quantity $d\ln\rho_{\rm DS}/d\ln f+d\ln C_\Lambda/d\ln f$ becoming so large that the $D$-function turns negative (see Eq.~\eqref{Eq.D}). As the inspiral proceeds, the density and speed of sound both rise such that the wake is only slightly supersonic $\mathcal{M}\gtrsim1$ and the $D$-function is positive once more.

\subsection{Bosonic vs fermionic equation of state}
Here we consider the full hydrostatic system given by Eqs.~\eqref{eq:structure_general} for the bosonic~\eqref{eq:EoSBS} and fermionic~\eqref{eq:EoSFS} EoSs. In this case, it is not possible to obtain a closed-form solution of the density profiles, so we calculate numerical solutions. The approximate expression for the $D$-function in the subsonic limit, Eq.~\eqref{eq:D_function_barotropic}, is still useful for interpreting the low-frequency behavior, but the numerical results shown below are obtained from the full expression in Eq.~\eqref{Eq.D} together with the numerically evaluated Coulomb logarithm $C_\Lambda$.
\begin{figure}[htbp]
\centering 
\includegraphics[width=0.48\textwidth]{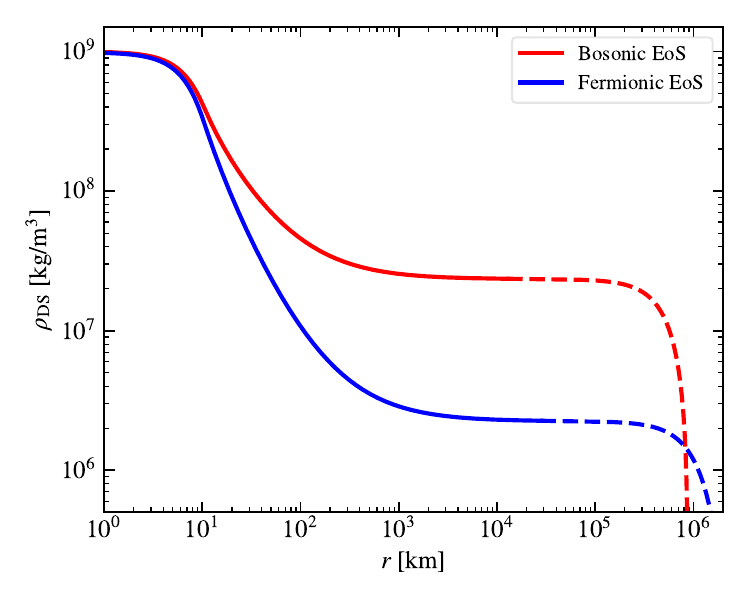} 
\includegraphics[width=0.48\textwidth]{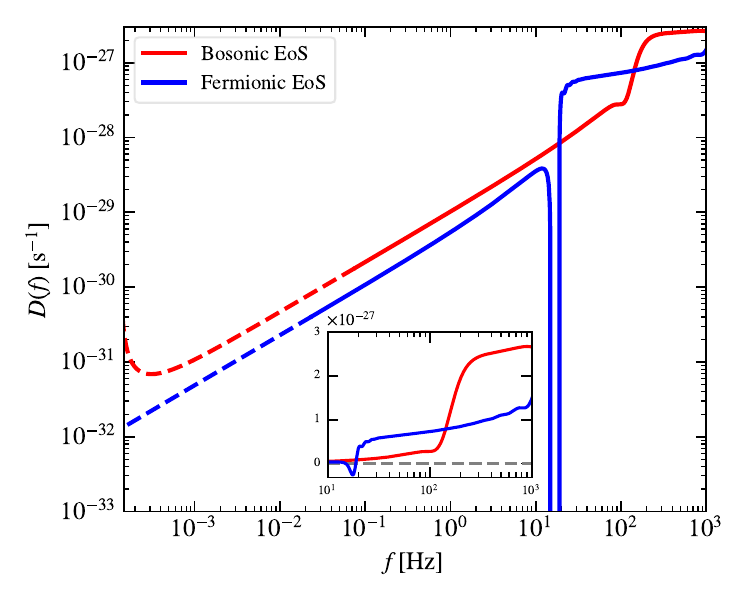} 
\includegraphics[width=0.48\textwidth]{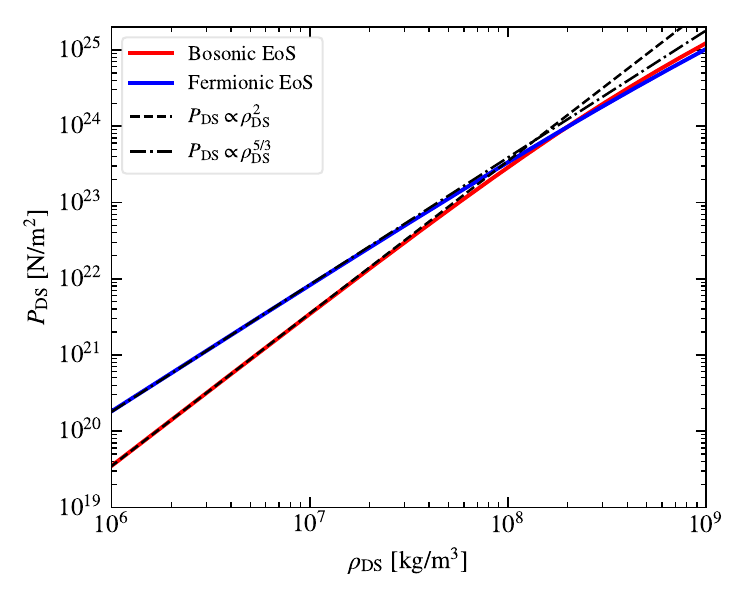}
\includegraphics[width=0.48\textwidth]{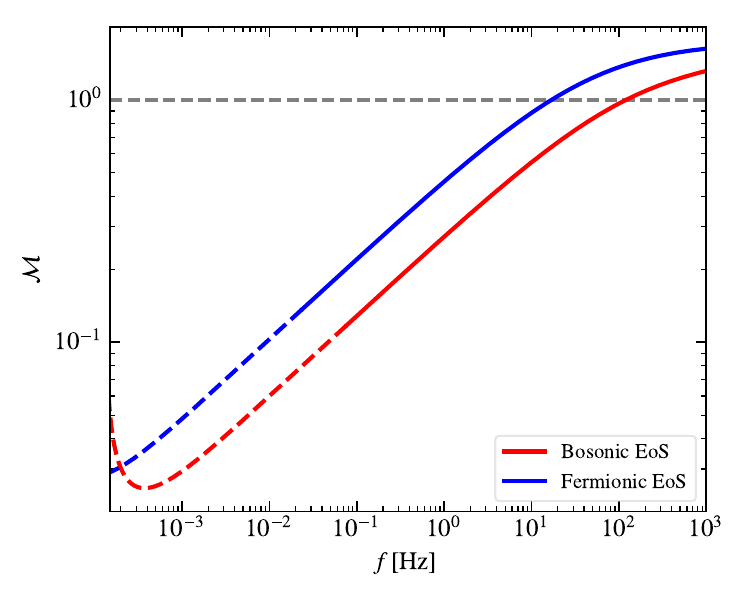}
\caption{\textit{Upper left panel:} Example DS density profiles for the bosonic~\eqref{eq:EoSBS} and fermionic~\eqref{eq:EoSFS} EoSs. \textit{Upper right panel:} $D$-function for the configurations in the left panel. \textit{Bottom left panel:} The EoSs in the density regions considered. The dashed and dash-dotted curves represent the polytropic relations expected to hold at small densities for bosons and fermions, respectively. \textit{Bottom right panel:} Mach number of the perturber as a function of frequency for both configurations. In all panels the total DS mass is fixed to $10^4M_\odot$ and the core density is $\rho_{\rm DS}=10^9~\mathrm{kg}/\mathrm{m}^3$. For concreteness, we set the NS core density to $\rho_{\rm NS,c}=7.5\times10^{17}~\mathrm{kg}/\mathrm{m}^3$, yielding a NS mass and radius of $M_{\rm NS}\sim 1\, M_\odot$ and $R_{\rm NS}\sim11.8~\rm km$. Additionally, in the right panels we consider a companion mass of $M_*=1\,M_\odot$. The colored dashed segments represent the region where the DS mass is larger than $0.1\,M_\odot$ and the approximate Kepler formula is no longer valid. We set the redshift of DS-NS system as $0.5$ in calculation.} 
\label{fig:BSvsFS}
\end{figure}

In Fig.~\ref{fig:BSvsFS}, we present a direct comparison between bosonic and fermionic DS configurations with the same total DS mass and the same central DM density. The bottom-left panel shows the corresponding equations of state. In the low-density limit, the bosonic EoS~\eqref{eq:EoSBS} behaves as $P_{\rm DS}\propto\rho_{\rm DS}^2$, while the fermionic EoS~\eqref{eq:EoSFS} behaves as $P_{\rm DS}\propto\rho_{\rm DS}^{5/3}$. Bosons therefore have a larger polytropic exponent ($2>5/3$) than fermions in this regime. At higher densities, the full expressions in Eqs.~\eqref{eq:EoSBS} and~\eqref{eq:EoSFS} deviate from these polytropic forms, and the nonlinear behavior of the two EoSs becomes apparent, as can be seen by comparing with the respective dashed and dash-dotted curves.

This difference in polytropic exponent is reflected in the density profiles shown in the upper-left panel. Because the one of the bosonic EoS is larger, the bosonic configuration is more compact; namely for fixed central density and total DS mass, it has a smaller radius than the fermionic configuration. The softer fermionic EoS produces a more extended configuration; consequently, at a fixed radius away from the center, the fermionic density is generally lower than the bosonic density. Because of this, at any fixed radius in the region where both profiles are nonzero, the fermionic density is lower than the bosonic density. Consequently, the enclosed DS mass remains subdominant over a larger range in radius, and so the Keplerian approximation remains valid to larger radii for the fermionic configuration. In all panels, the dashed colored segments in the curves indicate radii where the enclosed DS mass exceeds $0.1\,M_\odot$, corresponding to the regime in which the Keplerian approximation is no longer reliable.

The lower outer density and softer EoS of the fermionic configuration also reduce the sound speed, 
\begin{equation}
c_s\equiv\sqrt{\partial P_{\rm DS}/\partial \rho_{\rm DS}}~,
\label{Eq.Soundspeed}
\end{equation}
and therefore increase the Mach number of the perturber $\mathcal M$ defined in Eq.~\eqref{Eq.MachNo}, for a given orbital velocity. This behavior is visible in the bottom-right panel, where the fermionic Mach number grows more rapidly with frequency than the bosonic one. As a result, the fermionic configuration enters the transonic regime at lower frequencies and becomes supersonic earlier than the bosonic configuration.

This transition leaves a characteristic imprint on the $D$-function shown in the upper-right panel. Once the companion enters the supersonic regime, the derivative combination $d\ln\rho_{\rm DS}/d\ln f+d\ln C_\Lambda/d\ln f$ can exceed $11/3$, making the $D$-function in Eq.~\eqref{Eq.D} negative. For the fermionic configuration, this occurs around $f\sim20~\mathrm{Hz}$. On the other hand, for the bosonic configuration, the density and sound speed remain larger in the relevant region, so the Mach number stays smaller and the $D$-function remains positive over the displayed frequency range. Nevertheless, the bosonic configuration also eventually approaches the transonic regime, but only at larger frequencies, $f\sim10^2~\mathrm{Hz}$, as can be seen from the Mach number in the bottom-right panel.

Finally, at low frequencies, the orbital radius lies in the outer part of the DS, where the density profiles are approximately constant. Therefore $d\ln\rho/d\ln f\simeq0$, and the contribution of the density gradient to Eq.~\eqref{eq:D_func_approx} becomes subdominant. The $D$-function then approaches the simple scaling $D\propto f^{2/3}$, which is largely independent of the detailed EoS. Such behavior is clearly visible in the upper-right panel for both the bosonic and fermionic solutions.

\section{Reconstruction of the Dark Matter Equation of State}
\label{Sect.Cloud_Properties}

As discussed above, for a given DM EoS, the hydrostatic equations determine the equilibrium density profile of the dark component, while the derivative of the EoS determines the local sound speed $c_s$. The density and sound speed profiles then determine the local DF response through the Mach number $\mathcal{M}$, the Coulomb logarithm  $C_\Lambda$, and the corresponding DF power loss. Consequently, both the 
amplitude and frequency dependence of $D(f)$ encode information about the  underlying DM EoS.

In this section, we address the inverse problem: given an observed $D(f)$, can the underlying DM EoS be reconstructed without assuming a specific DM model? In the following sections, we will introduce the reconstruction method to achieve this target, and show two examples of the reconstruction with mock EoS for both boson star and fermion star. 

\subsection{Reconstruction method}

We formulate the reconstruction as a constrained optimization
problem. We initially assume a $P-\rho$ relation in the power-law form 
\begin{equation}
    P =  P_p \times \left(\frac{\rho}{\rho_p}\right)^{\Gamma_g} ~ ,   
\end{equation} 
where $P_p$ and $\rho_p$ are the reference pressure and density, respectively. This power-law EoS is used only to initialize the reconstruction and is not imposed on the final reconstructed EoS. Starting from this initial EoS, we allow its density-dependent slope to deform through a finite set of physically constrained control variables: 
\begin{equation}
    \boldsymbol{\theta}
    =
    \left(
        \theta_1,\ldots,\theta_{N_k},\theta_c
    \right),
    \qquad N_k=12 .
\end{equation}
The first \(N_k\) components control the density dependence of the
EoS slope $c_s^2(\rho_{\rm DS})$ at geometrically spaced control densities, while $\theta_c$
controls the central DM pressure. These variables do not represent independently varied pressure values. Instead, the EoS controls are interpolated over the logarithmic density grid and mapped through a bounded logistic transformation, such that
\begin{equation}
    0
    <
    \frac{\partial P_{\rm DS}}{\partial \rho_{\rm DS}}
    <
    0.98c^2
\end{equation}
for every trial model. The full pressure-density relation is then obtained by integrating the resulting slope profile subject to
$P_{\rm DS}(\rho_{\rm DS}=0)=0$. 
The final control variable determines $P_{\rm DS,c}$ through a separate bounded transformation satisfying
\begin{equation}
    P_{\rm DS}(\rho_1)
    \leq
    P_{\rm DS,c}
    \leq
    3P_{\rm DS}(\rho_{\max}) .
\end{equation}
Consequently, each 13-dimensional control vector
$\boldsymbol{\theta}$ uniquely generates a tabulated trial EoS and its central pressure while preserving monotonicity, stability, and subluminal sound propagation by construction.

Using the method introduced in Sec.~\ref{Sect.Density_Profile}, we can derive the density profile of the DS with guessed EoS, and then calculate the corresponding $D_g(f)$ using Eq.~\eqref{Eq.D} together with the Coulomb logarithm introduced in Sec.~\ref{Sect.CLambda}. The reconstructed parameters are obtained together with the target $D_t(f)$ curve obtained from the observation by minimizing
\begin{equation}
    \boldsymbol{\theta}_{\rm fit}
    =
    \underset{\boldsymbol{\theta}}{\operatorname{argmin}}\,
    L
    \left[
        D_g(f;\boldsymbol{\theta}),
        D_t(f)
    \right] ~ , 
    \label{Eq.fitting}
\end{equation}
where the total loss function $L$ is defined as 
\begin{equation}
    L(D_g,D_t) = L_+ + w_0 L_0 + w_e L_e + w_i L_i ~ .  
\end{equation}
Here $w_0$, $w_e$, and $w_i$ are the weights of different loss functions, which are defined as follows \cite{2023arXiv230105579C}: 
\begin{itemize}
\item $L_+$ is the loss function for the mismatch of $D_g(f)$ and $D_t(f)$, defined as 
\begin{equation}
    L_+ = \frac{1}{N_+} \sum_{i\in  \mathcal{I_+}}\left[\ln D_g(f_i) - \ln D_t(f_i) \right]^2 ~ .
\end{equation}
Here $\mathcal I_+ \equiv \{i|D_t(f_i)>0\}$ are the points where we have the positive target signal, and $N_+$ is the number of target data inside $\mathcal I_+$.  The logarithmic residual measures multiplicative rather than absolute deviations, preventing the fit from being dominated by the largest-amplitude part of a signal that spans several orders of magnitude. It also treats reciprocal multiplicative overestimates and underestimates symmetrically. For small deviations, this loss reduces approximately to the mean squared fractional error.

\item $L_0$ is the one-sided upper limit loss, which is defined as 
\begin{equation}
L_0 = \frac{1}{N_0} \sum_{i\in\mathcal I_0} \left[\max\left(0, \ln\frac{D_g}{D_{\rm upper}}\right)\right]^2 ~ , 
\end{equation} 
with $\mathcal I_0 = \{i| D_t(f_i) = 0 \}$. In the low frequency region, the target signal satisfies $D_t(f_i) = 0$, such that the logarithmic residual $\ln(D_g/D_t)$ is not well-defined. Instead, we impose an upper limit $D_{\rm upper} = 0.1 \min\{D_t(f_i)| D_t(f_i) > 0\}$ where the minimum is taken over all nonzero target values on the reconstructed signal, which is determined by the sensitivity of the GW detector. When the reconstructed signal exceeds this upper limit, we penalize the objective function with $L_0$, which prevents the optimization from producing a significant positive signal in the physically vanishing region, while avoiding the unnecessary requirement that an already negligible prediction be driven to numerical zero.

\item $L_e$ is the edge loss, which is defined as 
\begin{equation}
    L_e = \left[\ln\left(\frac{f_{\rm edge, g}}{f_{\rm edge,  t}}\right)\right]^2 ~ , 
\end{equation}
with $f_{\rm edge, g}$ and $f_{\rm edge, t}$ the reconstructed and target frequencies at which the orbital radius of the companion coincides with the boundary of DS
\begin{equation}
    r(f_{\rm edge}) = R_{\rm DS} ~ .
\end{equation}
The logarithmic ratio gives a dimensionless measure of the mismatch between the reconstructed and target DS radius. 

\item $L_i$ is invalid-density loss, which is defined as
\begin{equation}
    L_i = \frac{1}{N_+} \sum_{i\in \mathcal I_+} I_i^{\rm invalid} ~ , 
\end{equation}
with $I_i^{\rm invalid}$ 
an indicator function defined as 
\begin{equation}
I_i^{\mathrm{invalid}}
=
\begin{cases}
1, & \text{if the forward evaluation is invalid} ~ , \\[4pt]
0, & \text{if the evaluation is successful and acceptable} ~ .
\end{cases}
\end{equation}
It takes the value $1$ if the solution of the guessed model is non-physical or numerically invalid, and $0$ otherwise. The non-physical solution can be caused by several reasons, such as superluminal sound speeds, a density profile outside the prescribed domain, or a failure of the forward solver. 
\end{itemize}

\begin{figure*}[t]
    \centering

    \resizebox{\linewidth}{!}{%
    \begin{tikzpicture}[
        font=\small,
        text=textdark,
        fblock/.style={
            draw=forwardline,
            fill=white,
            line width=0.8pt,
            rounded corners=3pt,
            align=center,
            minimum height=13mm,
            text width=27mm,
            inner sep=4pt
        },
        fwide/.style={
            fblock,
            text width=35mm
        },
        iblock/.style={
            draw=inverseline,
            fill=white,
            line width=0.8pt,
            rounded corners=3pt,
            align=center,
            minimum height=13mm,
            text width=27mm,
            inner sep=4pt
        },
        iwide/.style={
            iblock,
            text width=35mm
        },
        forwardarrow/.style={
            -{Stealth[length=2.6mm,width=1.7mm]},
            draw=forwardline,
            line width=1.0pt
        },
        inversearrow/.style={
            -{Stealth[length=2.6mm,width=1.7mm]},
            draw=inverseline,
            line width=1.0pt
        },
        feedbackarrow/.style={
            inversearrow,
            dashed
        },
        paneltitle/.style={
            font=\bfseries\small
        }
    ]


    \node[fblock] (eos)
    {Guess $P(\rho)$};

    \node[fblock, right=6mm of eos] (tov)
    {Two-fluid\\
     TOV equations};

    \node[fwide, right=6mm of tov] (structure)
    {Hydrostatic structure\\[-1mm]
     $\rho(r),\,m(r),\,R_{\rm DS}$\\[0mm]
     $c_s^2$};

    \node[fwide, right=6mm of structure] (drag)
    {Orbital response\\[0mm]
     $\mathcal{M},\quad C_\Lambda$\\[0mm]
     $P_{\rm DF}(r)$};

    \node[fblock, right=6mm of drag] (dmodel)
    {$D_{g}(f)$};

    \draw[forwardarrow] (eos)       -- (tov);
    \draw[forwardarrow] (tov)       -- (structure);
    \draw[forwardarrow] (structure) -- (drag);
    \draw[forwardarrow] (drag)      -- (dmodel);

    \node[
        paneltitle,
        text=forwardline,
        above=4mm of eos.north west,
        anchor=south west
    ] (forwardtitle)
    {Forward model};


    \node[iblock, below=19mm of dmodel] (target)
    {$D_{t}(f)$};

    \node[iwide, left=8mm of target] (loss)
    {Objective function\\[0mm]
     $L
       (D_g,D_t)$};

    \node[iwide, left=8mm of loss] (optimizer)
    {Physically constrained\\
     optimization gets $\boldsymbol{\theta}_{\rm fit}$};

    \node[iblock, left=8mm of optimizer] (reconstructed)
    {$P_{\rm rec}(\rho_{\rm DS})$};

    \draw[inversearrow]
        (target.west) -- (loss.east);

    \draw[inversearrow]
        (loss.west) -- (optimizer.east);

    \draw[inversearrow]
        (optimizer.west) -- (reconstructed.east);

    \draw[inversearrow]
        (dmodel.south)
        -- ++(0,-6mm)
        -| (loss.north);

    \draw[feedbackarrow]
        (reconstructed.north)
        -- ++(0,2mm)
        -| node[
            pos=0.72,
            left,
            font=\scriptsize,
            text=inverseline
        ] {Update trial EoS}
        (eos.south);

    \node[
        paneltitle,
        text=inverseline,
        above=4mm of reconstructed.north west,
        anchor=south west
    ] (inversetitle)
    {Inverse reconstruction};


    \begin{scope}[on background layer]

        \node[
            fit={
                (forwardtitle)
                (eos)
                (tov)
                (structure)
                (drag)
                (dmodel)
            },
            fill=forwardblue,
            draw=forwardline!55,
            line width=0.7pt,
            rounded corners=5pt,
            inner xsep=5mm,
            inner ysep=4mm
        ] {};

        \node[
            fit={
                (inversetitle)
                (reconstructed)
                (optimizer)
                (loss)
                (target)
            },
            fill=inversegreen,
            draw=inverseline!55,
            line width=0.7pt,
            rounded corners=5pt,
            inner xsep=5mm,
            inner ysep=4mm
        ] {};

    \end{scope}

    \end{tikzpicture}%
    }

    \caption{Forward and inverse reconstruction pipeline.
    A trial DM EoS determines the
    hydrostatic structure through the two-fluid TOV equations.
    The resulting density and sound-speed profiles determine the
    orbital Mach number, the DF response, and the
    observable \(D_g(f)\).
    The reconstructed equation of state is obtained by minimizing
    the mismatch between \(D_g(f)\) and
    \(D_t(f)\) within the physically admissible EoS space.
    }
    \label{Fig_Pipeline}

\end{figure*}
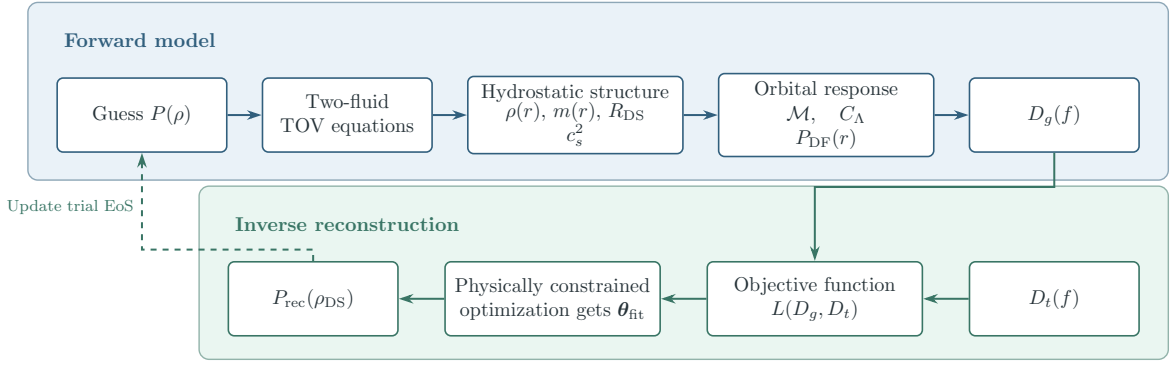

The minimization mentioned in Eq.~\eqref{Eq.fitting} is performed as follows: In each iteration, every control variable is first perturbed individually in the positive and negative directions with a perturbation length $\epsilon$ while all other variables are kept fixed. The corresponding forward models are evaluated to estimate each component of the loss gradient by finite differences 
\begin{equation}
    \frac{\partial L}{\partial\theta_i} \simeq \frac{L(\boldsymbol \theta + \epsilon \boldsymbol e_i)-L(\boldsymbol \theta - \epsilon \boldsymbol e_i)}{2\epsilon} ~ . 
\end{equation}
The resulting derivatives form the 13-dimensional gradient $\nabla_{\boldsymbol \theta} L$. A normalized descent direction is then defined as 
\begin{equation}
    \boldsymbol p = -\frac{\nabla_{\boldsymbol \theta} L }{{\rm max}_a |\partial L/\partial \theta_a|} ~ ,     
\end{equation}
and all control variables are updated simultaneously according to 
\begin{equation}
    \boldsymbol \theta^{(n+1)} = \boldsymbol \theta^{(n)} + \alpha \boldsymbol p ~ .     
\end{equation}
The step size $\alpha$ is selected by the following process: For each candidate step, the EoS and stellar structure are reconstructed and a new $D_g(f)$ and total loss $L$ are evaluated. The step size is then reduced until a sufficiently small loss is obtained. The accepted control vector is then used as the starting point for the next iteration, and the procedure is repeated until convergence of the loss function.

\subsection{Reconstruction examples}

In this section, we present the full reconstruction procedure and examine how well the reconstructed results reproduce the target observables. We begin with the mock EoS of boson and fermion shown in the lower left panel of Fig.~\ref{fig:BSvsFS}, with which we can construct the corresponding $D_t(f)$ diagram. Then, we apply our reconstruction method to recover the EoS from the target $D_t(f)$ diagram. 

\begin{figure}[htbp]
\centering 
\includegraphics[width=0.49\textwidth]{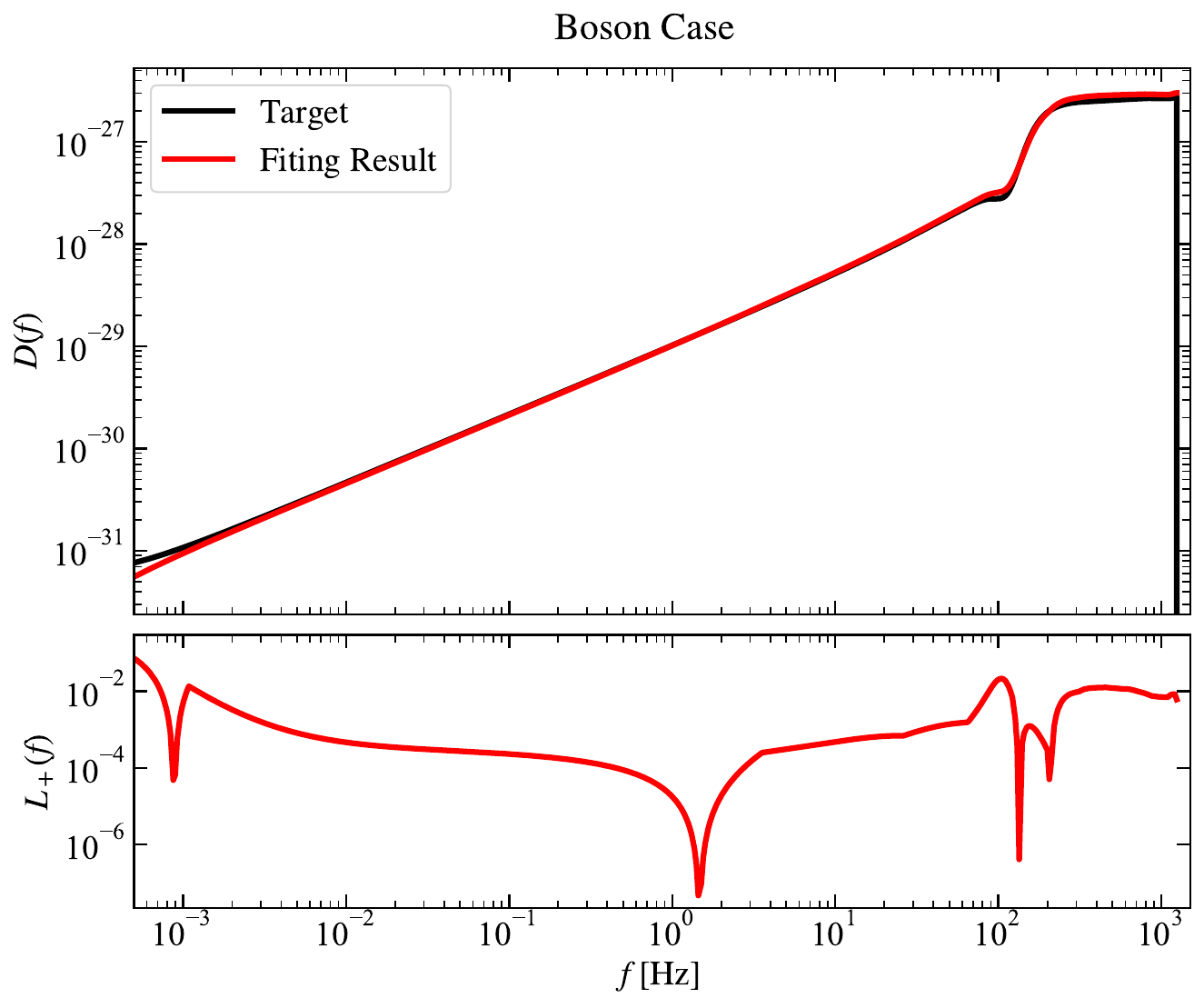} 
\includegraphics[width=0.49\textwidth]{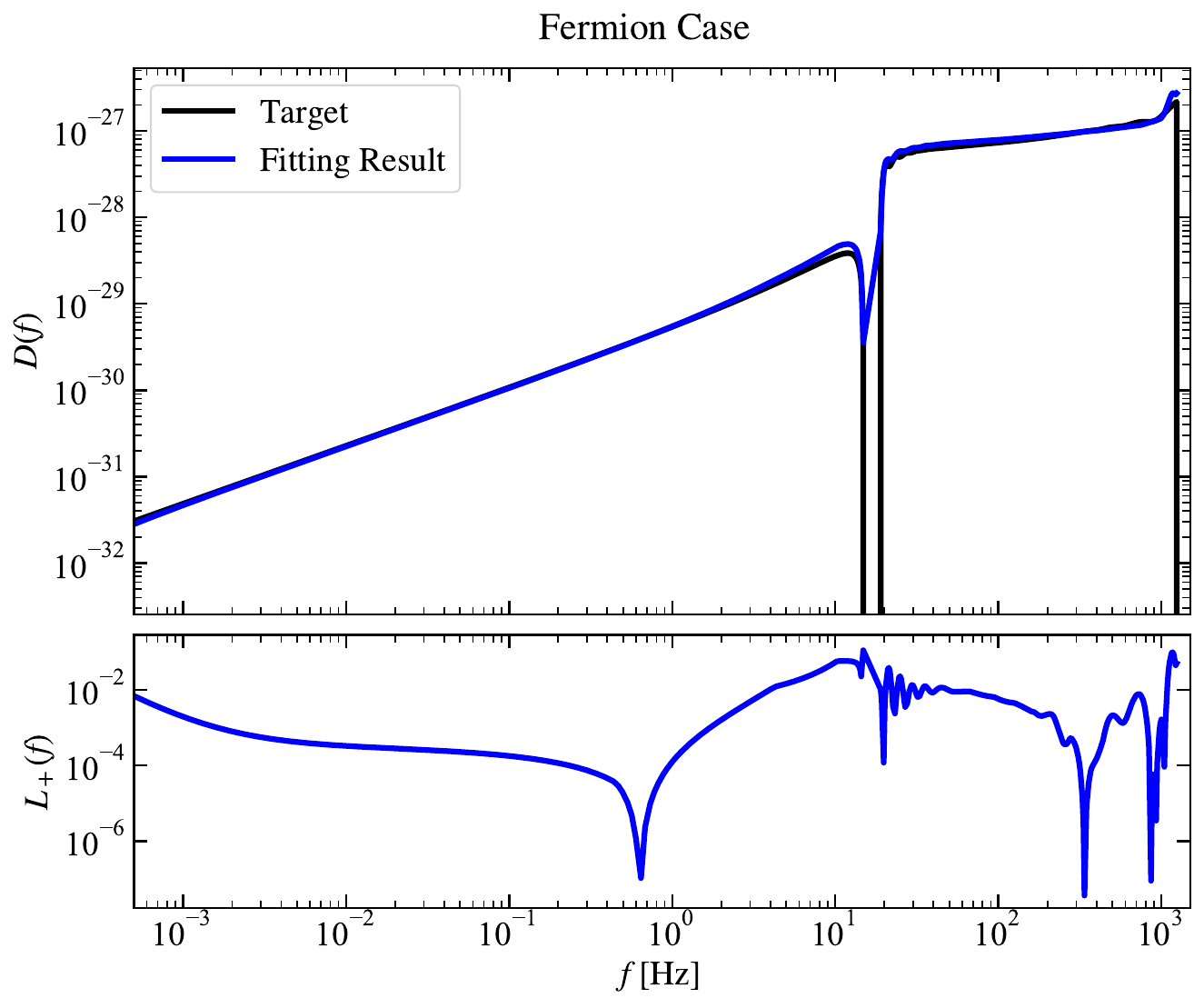} 
\includegraphics[width=0.49\textwidth]{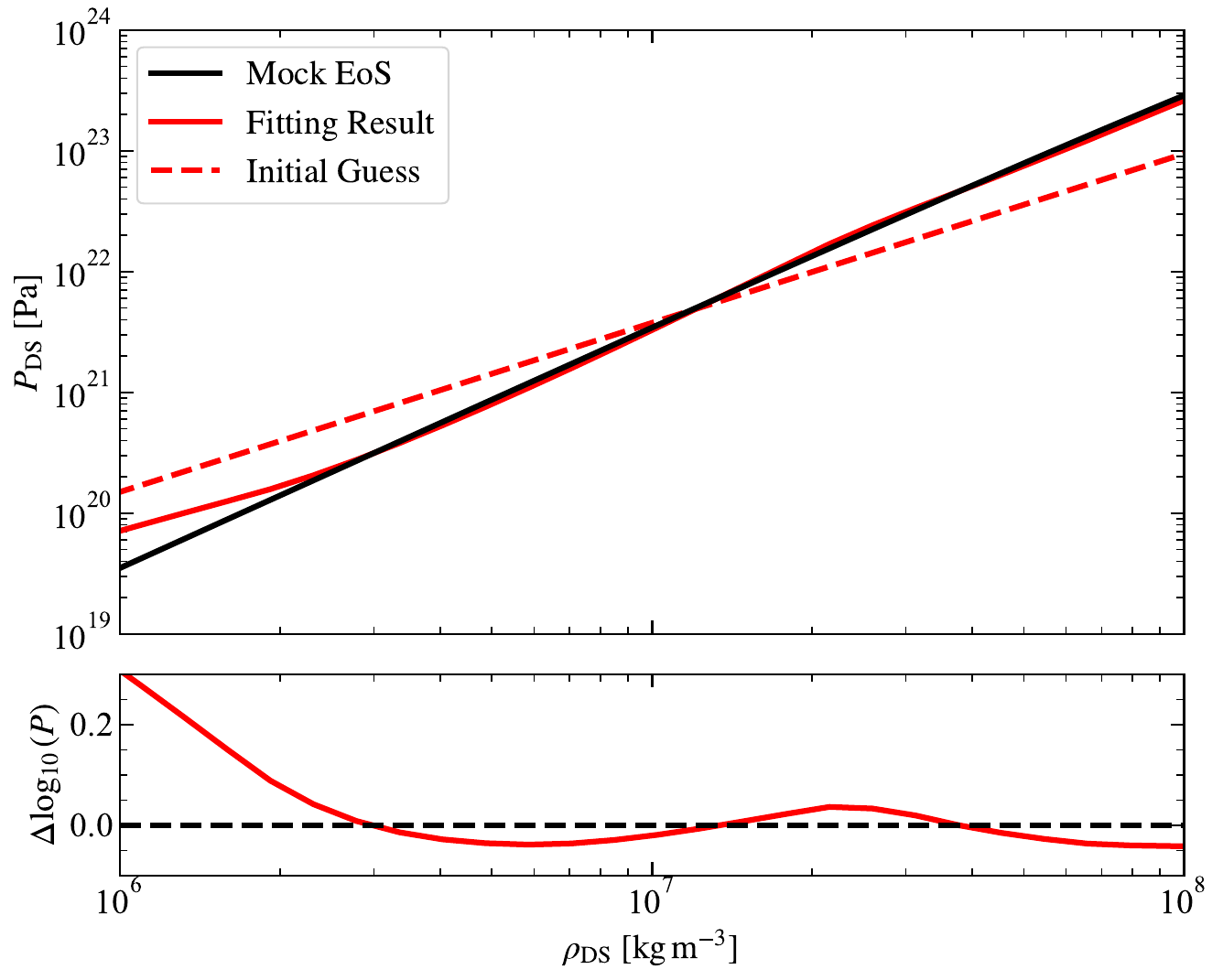} 
\includegraphics[width=0.49\textwidth]{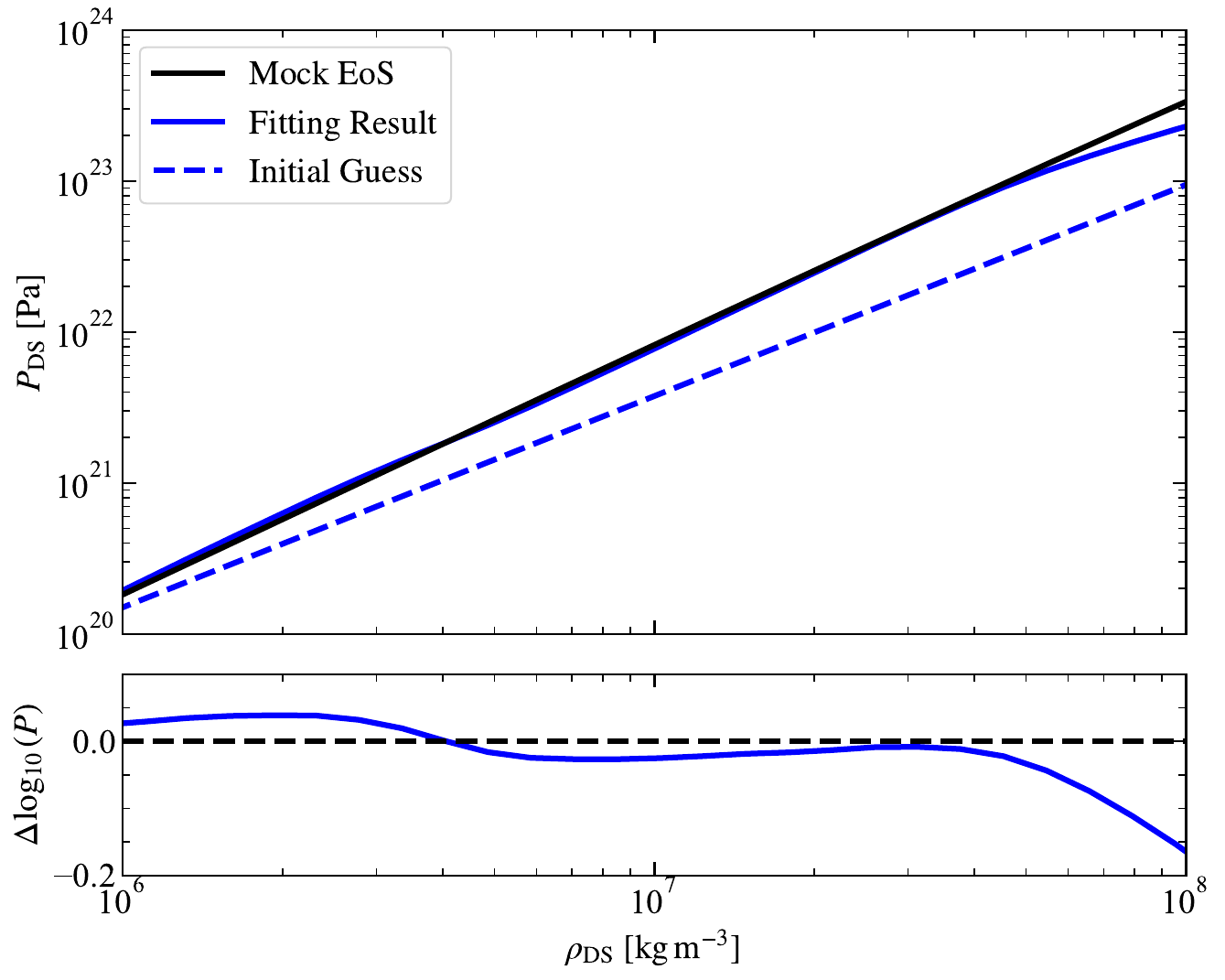} 
\caption{Reconstruction of the observable $D(f)$ and the corresponding positive loss $L_+(f)$ for both the bosonic and fermionic mock data.
\textit{Upper panels:} Comparison between the target and reconstructed $D(f)$ signals for the bosonic case (left) and the fermionic case (right), together with the corresponding positive loss $L_+(f)$. The black curves denote the mock $D(f)$ signals constructed from the bosonic and fermionic mock EoSs shown in the lower-left panel of Fig.~\ref{fig:BSvsFS}, while the red and blue curves show the reconstructed results for the bosonic and fermionic cases, respectively.
\textit{Lower panels:} Reconstruction of the EoS for the bosonic case (left) and the fermionic case (right), together with the corresponding reconstruction error $\Delta\log_{10}(P)$ defined in Eq.~\eqref{Eq_FittingError}. In both cases, we start from the same initial EoS with the parameter choice
$\{P_p = 1.5 \times 10^{20}\,{\rm Pa},\, \rho_p = 10^6\,{\rm kg\,m^{-3}},\, \Gamma_g = 1.4\}$.
The black curves denote the mock EoSs, while the solid red and blue curves show the reconstructed EoSs for the bosonic and fermionic cases, respectively. The dashed red and blue curves denote the corresponding initial EoS guesses. The bottom panels show the reconstruction errors for the two cases.} 
\label{fig:FittingDf}
\end{figure}

Starting with the same initial EoS for both bosonic and fermionic reconstruction 
\begin{equation}
P = P_p \times \left(\frac{\rho}{\rho_p}\right)^{\Gamma_g} ~ , 
\end{equation}
with the choice of parameters $\{ P_p = 1.5 \times 10^{20} \, {\rm Pa}, \, \rho_p = 10^6 \, {\rm kg/m^3}, \, \Gamma_g = 1.4\}$, the corresponding fitting results for the $D-f$ diagram are shown in Fig.~\ref{fig:FittingDf}. After the reconstruction, the positivity loss associated with the $D$-function eventually stabilizes at the order of $L_+\sim10^{-2}$, at which point we terminate the regression procedure. In the lower two panels we demonstrate the corresponding fitting result for EoS, with the fitting error defined as 
\begin{equation}
    \Delta\log_{10}(P) \equiv \log_{10} P_{\rm rec} - \log_{10} P_{\rm mock} ~ . 
    \label{Eq_FittingError}
\end{equation}

As shown in Fig.~\ref{fig:FittingDf}, the reconstruction accuracy is not uniform over the full density range. For both the bosonic and fermionic cases, the reconstructed $D_g(f)$ reproduces the target $D_t(f)$ well over most of the relevant frequency interval, and the corresponding EoS can be recovered 
with an accuracy of $|\Delta\log_{10}(P)| \lesssim 0.05$ over the best-fitting density range. The reconstruction is generally more accurate in the lower and intermediate density regions, where the companion moves in the subsonic regime and the $D$-function has a relatively simple and smooth dependence on the density profile and sound speed, allowing the reconstruction to constrain the EoS efficiently.

The accuracy decreases around the edges of the density interval, which is particularly visible in the fermionic case at high densities. As the companion moves inward, the smaller sound speed of the fermionic configuration causes the orbit to enter the transonic and supersonic regimes earlier, where the $D$-function develops sharp frequency dependent features and can change sign. The points with $D_t(f)\leq 0$ are not directly included in the present logarithmic loss function, and therefore provide weaker constraints on the corresponding local EoS. At higher densities, the orbital separation approaches the radius of the central NS. For $\rho_{\rm DS}\sim10^8\,{\rm kg\,m^{-3}}$ in the fermionic example, the separation is already of order $\sim10\,{\rm km}$. The system is then close to the end of the inspiral, where tidal effects, nonlinear dynamics, and eventually merger become important, and the assumptions underlying our reconstruction are no longer expected to remain reliable.

These results emphasize that the reconstruction is constrained only by the portion of the DS profile that is both sampled by the inspiral and reliably 
encoded in the observable $D(f)$, rather than by the entire EoS over all densities. The reconstructed $P_{\rm DS}(\rho_{\rm DS})$ should therefore be 
interpreted as a local inference over the density interval effectively probed by the GW signal. This point is important for assessing what physical 
information can be extracted from the reconstructed EoS, which we discuss in the following subsection.


\subsection{Physical interpretation of a reconstructed equation of state}\label{subsec:interpretation}
The reconstruction of the DM density profile and EoS from the $D$-function should be interpreted as a local measurement over the density interval actually probed by the inspiral. The low-frequency part is sensitive to the outer layers of the DS, while the high-frequency part would, in principle, probe the denser inner region. In practice, however, the usable frequency window is limited at high frequencies because, as the binary separation decreases, tidal effects, nonlinearities in the dark medium response and eventually merger dynamics become important, and our approximations cease to be reliable. Therefore, the highest densities reached in the DS are not necessarily accessible to this method. Outside the sampled range $[\rho_{\rm min},\rho_{\rm max}]$, the reconstructed $P_{\rm DS}(\rho_{\rm DS})$ is an extrapolation and should not be overinterpreted. Nevertheless, even this partial information can provide valuable insight into the underlying DM physics.

The first robust conclusion from a measured $D$-function is the presence of a dark environment is capable of exerting DF on the inspiraling companion. In the formalism we use in this paper, the additional power loss is determined by the local DM density and by the response of the medium through the Coulomb logarithm, as in Eq.~\eqref{eq:P_DF}. Therefore, within the dark-medium interpretation considered here, a non-zero $D$-function indicates an additional dissipative environment surrounding the binary. Since the particular frequency dependence of $D(f)$ depends on the speed of sound and density profile of that medium, the observation also shows that the environment has a well-defined pressure response, and, within the fluid description adopted here, implies a nontrivial pressure response characterized by a finite sound speed.

Once the EoS is reconstructed, its functional form can be used to distinguish between broad classes of DM models. As an example, if the reconstruction gives an approximate linear relation $P_{\rm DM}\simeq K\rho_{\rm DM}$, this would be consistent with a non-relativistic isothermal gas at constant temperature satisfying $P_{\rm DM}=(k_B T_{\rm DM}/m_\chi)\rho_{\rm DM}$. Under this assumption, the coefficient $K$ constrains the combination $T_{\rm DM}/m_\chi$. While an independent mass or temperature measurement or an additional assumption would be required to extract the quantities separately, such a result would already point towards pressure being dominated by thermal motion rather than by degeneracy or self-interactions. On the other hand, a quadratic relation $P_{\rm DM}\simeq K\rho_{\rm DM}^2$ is characteristic of several self-interacting DM models. As previously discussed, such behavior arises naturally for repulsively self-interacting bosons in the low-density limit and can also occur for self-interacting fermions in an intermediate-density regime where the interaction pressure dominates over degeneracy pressure. In this case, the coefficient $K$ also constrains combinations of microscopic parameters: for the bosonic model, $K$ is directly related to the combination $\lambda^{-1}m_\chi^4$, while for the fermionic model the coefficient depends on a combination of $\alpha$, $m_\chi$, and $m_\phi$. Therefore, a measurement of $K$ generally constrains a parameter combination rather than determining individual masses and couplings separately.

The reconstructed EoS could also provide information about the quantum nature of the DM. In particular, a relation of the form $P\propto\rho^{5/3}$ over an appropriate density range would be characteristic of non-relativistic degenerate fermionic matter, while $P\propto\rho^{4/3}$ would indicate the relativistic-degeneracy regime. Observation of these characteristic scalings would therefore provide strong evidence for quantum degeneracy and would favor a fermionic description of DM. More generally, it is useful to define the effective polytropic index $\Gamma_{\rm eff}\equiv d\ln P_{\rm DM}/d\ln\rho_{\rm DM}$. Its density dependence can reveal transitions between different physical contributions to the pressure. For example, in a self-interacting fermionic medium one could identify a transition from a degeneracy-dominated regime with $\Gamma_{\rm eff}\simeq5/3$ to an interaction-dominated regime with $\Gamma_{\rm eff}\simeq2$. The location of such a transition provides additional information about the relative importance of the Fermi pressure and the interaction pressure.

The EoS does not need to be a single power law. A sufficiently precise reconstruction of a nonlinear relation $P_{\rm DM}=P_{\rm DM}(\rho_{\rm DM})$ can be compared directly with predictions from microscopic models, such as an effective field theory with particular self-interaction operators or a model with a mediator of specified mass and coupling. In this way, EoS reconstruction relates the macroscopic structure of the DS and the microscopic particle physics responsible for its pressure support. Because different microscopic models can produce similar EoSs over a restricted density range, such comparisons usually constrain classes of models and combinations of parameters rather than providing a unique identification of the underlying theory. 

\section{Detectability of GWs from Compact Object Inspirals in Dark Stars} 
\label{Sect.ObservationalConstraints}

In this section, we discuss several conditions that must be satisfied for the DF signal to be detectable in GW observations and for our formalism to be valid.

First, in our work, we assume that the DS mass is much larger than the NS mass $M_{\rm NS}$, so that the DS is not significantly depleted during the capture of the NS. However, when considering the inspiral motion of the binary companion, if the enclosed DS mass is too large, Eq.~\eqref{Eq.dfdt} is no longer valid, and detected GW frequency $f$ should be modified as
\begin{equation}\label{eq:Kepler}
f=\frac{1}{\pi}\sqrt{\frac{G\left[M_{\rm NS}+m_{\rm DS}(r)+M_*\right]}{(1+z)^2r^3}} ~ .
\end{equation}
Here, $m_{\rm DS}(r)$ denotes the enclosed DS mass within the orbital radius $r$, and the relation between the GW frequency in the source frame $f_r$ and observer frame $f$ satisfies $f_r \equiv (1+z) f$. Therefore, to ensure the validity of 
$D$-function, we require the enclosed DM mass to be much smaller than the mass of the binary companion $M_*$, satisfying
\begin{equation}\label{eq:requirement_mass}
    m_{\rm DS}(r)/M_{\rm NS} < 0.1~.
\end{equation}

\begin{figure}[htbp]
\centering 
\includegraphics[width=0.48\textwidth]{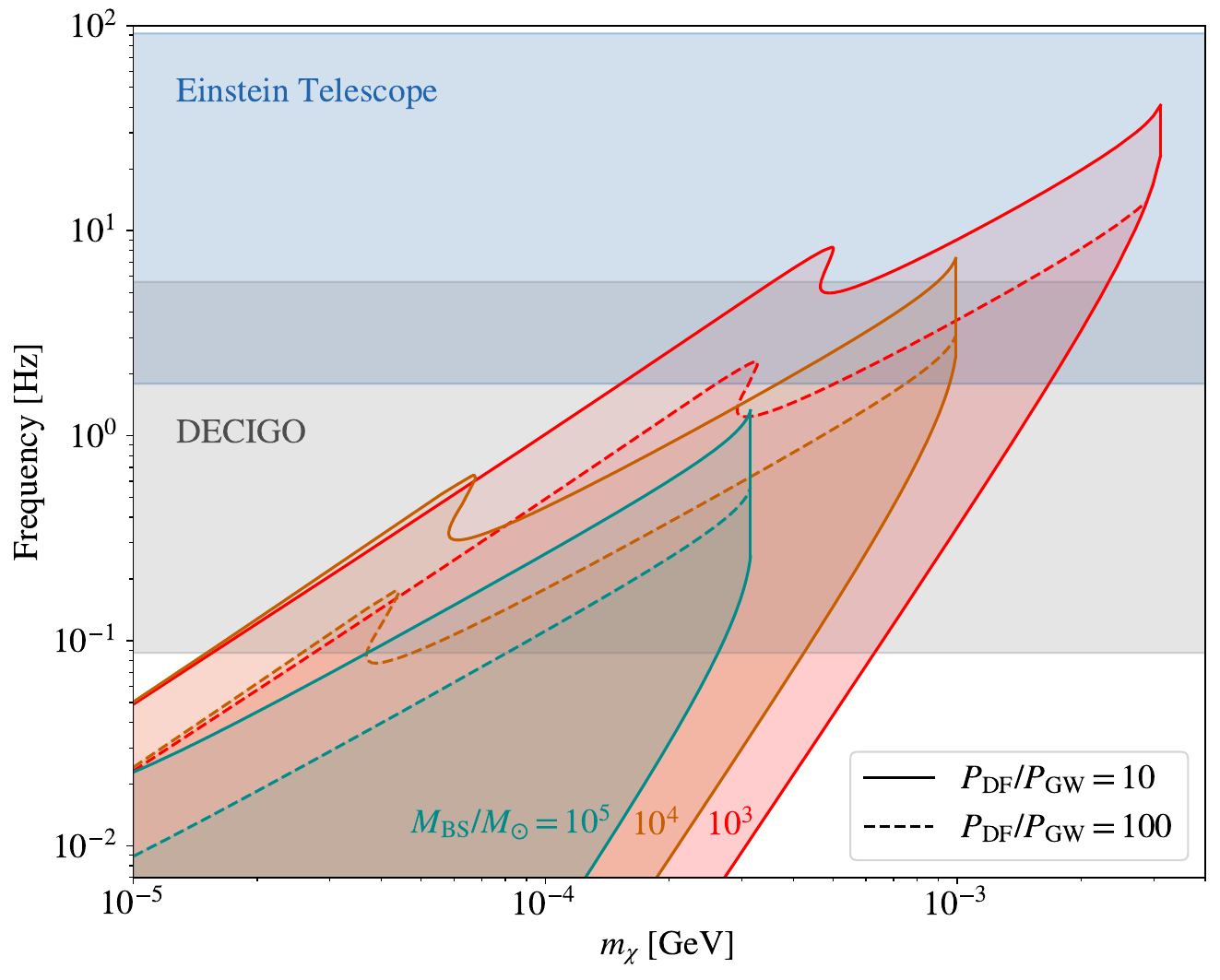}
\includegraphics[width=0.48\textwidth]{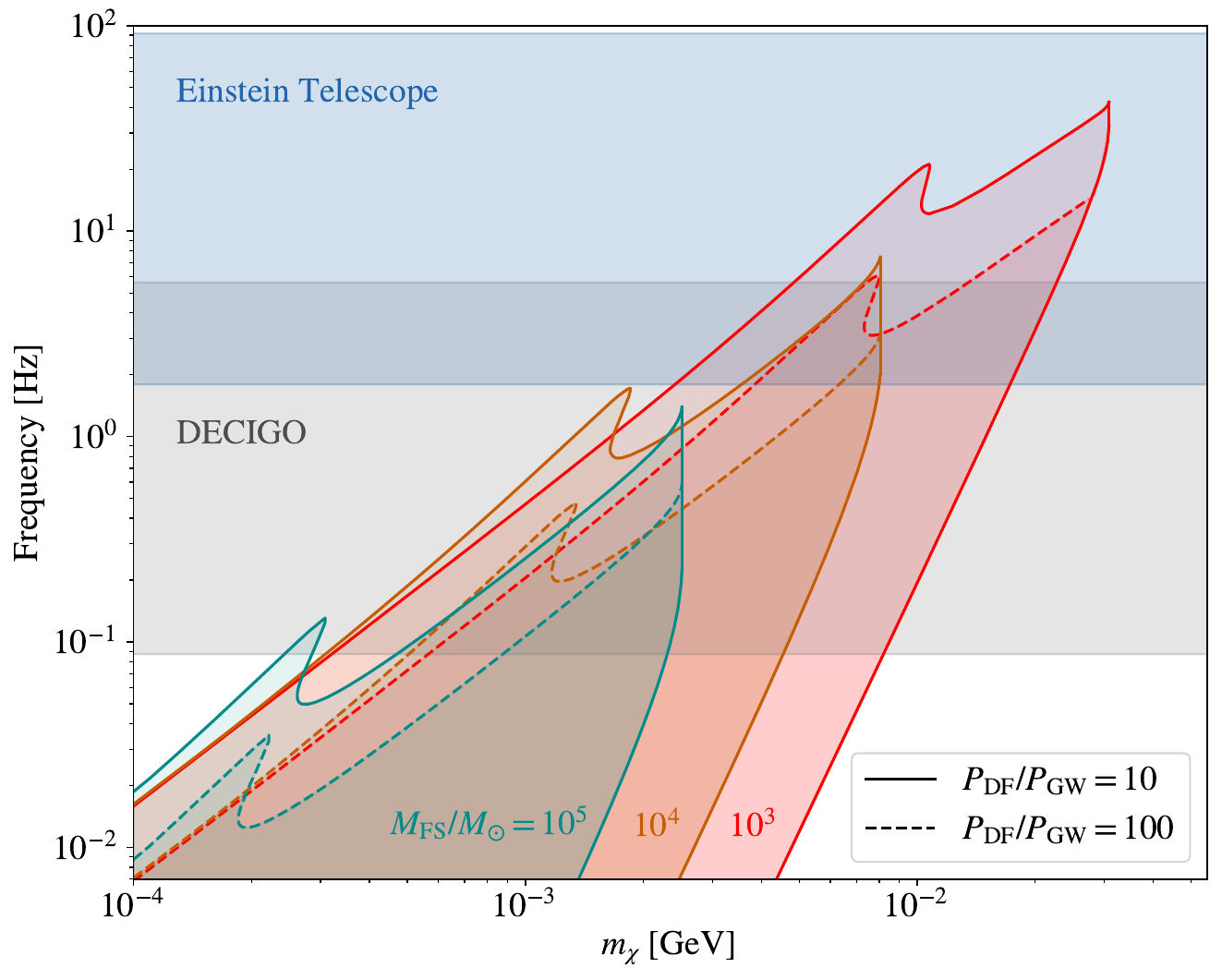}
\caption{\textit{Left panels:} The detectable GW frequency range with various DM mass in the boson star. The cyan, orange and red shadow regions correspond with DS mass $10^5, 10^4,10^3 \, M_\odot$, respectively, and they satisfy $m_{\rm DS}/M_{\rm NS}<0.1$ and $P_{\rm DF}/P_{\rm GW}> 100$ (dashed), $10$ (solid). The blue and gray shadow regions represent the detectable frequency ranges of Einstein Telescope and DECIGO. We consider the bosonic EoS with $\lambda=0.01$. \textit{Right panels:} The detectable GW frequency range with various DM mass in the fermion star. The cyan, orange and red shadow regions correspond with DS mass $10^5, 10^4,10^3 \, M_\odot$, respectively, and they satisfy $m_{\rm DS}/M_{\rm NS}<0.1$ and $P_{\rm DF}/P_{\rm GW}> 100$ (dashed), $10$ (solid). The blue and gray shadow regions represent the detectable frequency ranges of Einstein Telescope and DECIGO. We consider the fermionic EoS with $\alpha/m_\phi^2=10^{-3}~\mathrm{MeV}^{-2}$.} 
\label{fig:ObservationalConstraints}
\end{figure}

Second, we also require the dissipative power governing the evolution of the GW waveform to be dominated by DF, so that the corresponding signal is not buried by the standard GW contribution and background noise. This requires the DF power loss to be at least larger than the GW power loss, and we consider a conservative power ratio $10$ as a benchmark constraint (we also use a power ratio $100$ as a comparison in Fig.~\ref{fig:ObservationalConstraints}) as
\begin{equation}\label{eq:requirement_Power}
    P_{\rm DF}/P_{\rm GW}>10 ~ . 
\end{equation} 

The third requirement is that the GW signals from compact object inspirals in DS be detectable in the GW detectors such as Einstein Telescope and DECIGO. A GW signal detection requires its signal-to-noise ratio (SNR) is larger than a threshold value within the observation time. We follow Ref.~\cite{Rosado:2015voo, Ding:2020ykt}, the optimal SNR is defined as
\begin{equation}\label{eq:SNR}
    \mathrm{SNR} = \sqrt{4 \int_{f_{\rm min}}^{f_{\rm max}} \frac{|\tilde{h}(f)|^2}{S_n(f)} df}~,
\end{equation}
where $\tilde{h}(f)$ is the Fourier transform of GW waveform of a compact object, which is given by
\begin{equation}
    \tilde{h}(f) = \sqrt{\frac{5}{24}} \frac{(G \mathcal{M}_c (1+z))^{5/6}}{\pi^{2/3} c^{3/2} d_L(z)} f^{-7/6}~.
\end{equation}
In Eq.~\eqref{eq:SNR}, $f_{\rm min}$ and $f_{\rm max}$ are the initial and final observed GW frequencies during the observation period, which is taken to be $1 \, \mathrm{day}$ for Einstein Telescope and $1 \, \mathrm{year}$ for DECIGO. $S_n(f)$ is the detector's noise strain \cite{Moore:2014lga}. To ensure a detection probability of GW events greater than $95\%$ and a false-alarm probability below $0.1\%$, the optimal SNR must exceed the conservative threshold given in~\cite{LIGOScientific:2016vbw} as
\begin{equation}\label{eq:requirement_SNR}
    \mathrm{SNR} > 8~.
\end{equation}
The SNR criterion concerns detectability of the GW signal itself, while the $P_{\rm DF}/P_{\rm GW}$ criterion is introduced as a requirement for the environmental contribution due to DF to be distinguishable from the vacuum contribution.

It should be noticed that the above requirement on SNR is based on matching of GW templates \cite{Ajith:2007kx} and current templates are modeled for the compact binaries without the dark dense environment. Hence, the detection of GW with a $D$-function requires either GW templates for the compact binaries with the dark dense environment or template-independent methods, such as sine-Gaussian wavelets~\cite{Cornish:2014kda, LIGOScientific:2016fbo} or deep-learning neural networks~\cite{George:2017pmj}.

According to Eqs.~\eqref{eq:requirement_mass}, \eqref{eq:requirement_Power} and \eqref{eq:requirement_SNR}, we can obtain the detectable parameter regions in boson-star and fermion-star in Fig.~\ref{fig:ObservationalConstraints}.
We present the regions in the frequency-DM mass plane where Eqs.~\eqref{eq:requirement_mass}, \eqref{eq:requirement_Power} and \eqref{eq:requirement_SNR} are valid, such that the ratio of enclosed masses is always dominated by the NS, DF power dominates over GW power, and corresponding GW signals can be detected by Einstein Telescope and DECIGO. We consider three total DS masses: $M_{\rm BS}=10^3, 10^4$, and $10^5M_\odot$ and our two benchmark EoS (bosonic case in the left panel and fermionic case in the right panel). The upper limit on the DM mass at the right edge of both panels arises because, for larger DM particle masses, no DS configurations exist with total masses in the range $10^3-10^5M_\odot$. Meanwhile, we present the detectable frequencies in the future Einstein Telescope and DECIGO experiments, where we set observation time $1$ day for Einstein Telescope and $1$ year for DECIGO. In both panels we fix $\rho_{\rm NS,c}=7.5\times10^{17}~\mathrm{kg}/\mathrm{m}^3$, $z=0.5$, and $M_*=1\,M_\odot$. 

As a result, the lower and upper frequency bounds of the observationally relevant region are set primarily by the requirement that DF dominates over GW emission, rather than by the validity of the Keplerian approximation. In terms of prospects for experimental detection, this means that the spatial extent and compactness of the DS are less important than the DF power loss $P_{\rm DF}$.

\section{Discussions and Conclusions}
\label{Sect.DiscussionandConclusion}

In this work, we investigate the possibility of probing the EoS of DM  through the GW signal of a compact-object inspiral inside an extended DM configuration. We considered a NS embedded in a DS in hydrostatic equilibrium, with a compact companion inspiraling through the surrounding dark medium. In addition to the standard energy loss through GW emission, the companion experiences DF, which modifies its orbital evolution and therefore leaves an imprint of the DM properties on the GW signal. The main idea of this work is that this additional dissipation does not only probe the local DM density, it also carries information about the EoS of the medium.

To obtain physical intuition, we try to construct a simple model of the DM configuration, which is described by a polytropic EoS $P=K\rho^2$, which is relevant to the self-interacting DM models studied here. Under the assumption that the NS dominates the gravitational potential, the DM density profile can be solved analytically. In the subsonic regime, this also leads to an analytical approximation for $D(f)$.  In particular, when the inspiral probes an approximately constant-density region of the DM configuration, the observable follows the characteristic scaling $D(f)\propto f^{2/3}$. This result will be invalid when the density gradient becomes important or when the orbital motion approaches the sonic transition. 

We then formulated the inverse problem as a physical constrained optimization problem. Rather than assuming that the final EoS belongs to a fixed particle model or to a single polytropic family, we parameterized its density-dependent slope and reconstructed it by minimizing the mismatch between the predicted and target $D(f)$. 
Using the noiseless mock EoS data for bosonic and fermionic DSs, we find that starting from the same initial EoS guess, the optimization successfully reproduces both target $D(f)$ signal and recovers the underlying EoS with an accuracy of $|\Delta \log_{10}(P)| \lesssim 0.05$ over the best fitting density range. The reconstruction becomes increasingly robust toward higher densities, corresponding to smaller orbital separations where the DF signal is appreciable. By contrast, the low-density outer region is less well constrained because $D(f)$ becomes small there. In this regime the GW signal primarily determines the location of the boundary of DS rather than the detailed local EoS.

The applicability of the reconstruction has several physical constraints. In the regime studied here, the enclosed DM mass must remain sufficiently small that the orbital dynamics are approximately Keplerian and dominated by the compact objects, while the DS must remain extended beyond the neutron-star radius and satisfy causal sound-speed constraints. In addition, an observational reconstruction requires the DF contribution to be sufficiently large compared with the vacuum GW energy loss over the relevant frequency interval. These conditions define the region of DS parameter space in which the framework presented here can consistently be applied. 

Several extensions will be important before applying this method to realistic GW data. A full parameter-estimation analysis should incorporate detector noise and correlations between the environmental signal and the intrinsic binary parameters, rather than treating $D(f)$ as an exactly reconstructed observable. It will also be useful to relax the assumptions of circular and coplanar motion and to study regimes in which the enclosed DM mass, tidal response of the DS, or backreaction of the inspiraling companion can no longer be neglected.

\section*{Acknowledgments}

The authors would like to thank Masahide Yamaguchi for useful discussions. B.B.K, Q.D. and H.Y.Z. are supported by IBS under project code IBS-R018-D3.

\appendix

\section{Derivation of $D$--Function}\label{app:derivation_D}

Consider a compact binary in which one component is surrounded by a dense dark environment, such as an extended DS, DM spike, or superradiant boson cloud. The second compact object moves through this dark environment and experiences DF, with the corresponding power loss given by
\begin{equation}
    P_{\rm DF} = - 4 \pi \frac{G^2 M_*^2}{v} \rho C_\Lambda~,
\end{equation}
where $M_*$ is the mass of the compact binary component moving through the dark environment, $\rho$ is the density of the dark environment, $v$ is the relative velocity between the compact object and the dark environment, and $C_\Lambda$ is the Coulomb logarithm.

In the presence of the dark environment, the GW frequency in the source frame, $f_s$, evolves as
\begin{equation}\label{eq:frequency_evolution}
    \frac{d f_s}{d t} = - \frac{3}{(\pi G)^{2/3}} \frac{f_s^{1/3}}{\mathcal{M}_c^{5/3}} (P_{\rm GW} + P_{\rm DF})~,
\end{equation}
where $\mathcal{M}_c$ is the chirp mass of the compact binary and $P_{\rm GW}$ is the GW emission power. For a circular orbit, it can be expressed as
\begin{equation}
    P_{\rm GW} = - \frac{32}{5} \frac{G^{7/3}}{c^5} (\pi \mathcal{M}_c f_s)^{10/3}~.
\end{equation}
We next consider the GW frequency in the observer frame, $f$, which is related to the source-frame frequency by
\begin{equation}
    f_s = (1+z)f~,~~~dt_s = (1+z)^{-1}dt~.
\end{equation}
Here $z$ is the cosmological redshift of the compact binary, and $dt_s$ and $dt$ are time intervals measured in the source and observer frames, respectively. Transforming Eq.~\eqref{eq:frequency_evolution} to the observer frame gives
\begin{equation}\label{eq:GW_freq_obs}
    \frac{df}{dt} = \frac{96}{5}\frac{\left[G\mathcal M_c(1+z)\right]^{5/3}\pi^{8/3}f^{11/3}}{c^5}+\frac{3 f^{1/3}}{(\pi G)^{2/3}[\mathcal M_c(1+z)]^{5/3}}|P_{\rm DF}| ~.
\end{equation}
Here, the first term on the right hand side is the contribution from GW emission power and the second term on the right hand side is the contribution from DF power. In order to remove the contribution of GW emission in observed frequency evolution, we introduce the GW amplitude $h$ that reads
\begin{equation}\label{eq:GW_amp}
    h = \frac{4 \pi^{2/3}}{d_L(z)} \frac{[G \mathcal{M}_c (1+z)]^{5/3}}{c^4} f^{2/3}~,
\end{equation}
where $d_L(z)$ is the luminosity distance between the compact binary and the observer. Substituting Eq.~\eqref{eq:GW_amp} into Eq.~\eqref{eq:GW_freq_obs}, we define the time-dependent function $g(t)$ as
\begin{equation}
    g(t) \equiv \frac{1}{hf^3} \frac{df}{dt} = \frac{24}{5} \frac{d_L(z) \pi^2}{c} + \frac{12 G}{c^4 d_L(z)} \frac{|P_{\rm DF}|}{h^2 f^2}~.
\end{equation}
The function $g(t)$ consists of two terms: a time-independent term, $24\pi^2d_L/(5c)$, and a time-dependent term, $12G|P_{\rm DF}|/(c^4d_Lh^2f^2)$. In the absence of DF, $g(t)$ is therefore time independent. We can consequently take its time derivative to eliminate the GW contribution,
\begin{equation}\label{eq:dgdt}
    \frac{dg}{dt} = \frac{12 G}{c^4 d_L(z)} \frac{|P_{\rm DF}|}{h^2 f^3} \frac{df}{dt} \left(\frac{d \ln \rho}{d \ln f} + \frac{d \ln C_\Lambda}{d \ln f} - \frac{11}{3}\right)~.
\end{equation}
We now construct a quantity that depends only on the observables $h$, $f$, and $df/dt$. Multiplying Eq.~\eqref{eq:dgdt} by $-h^2f^3/(df/dt)$, we define the $D$-function as
\begin{align}\nonumber
        D &\equiv -h^2f^3 \frac{dg/dt}{df/dt} = \frac{dh}{dt} + 3 \frac{h}{f} \frac{df}{dt} - \frac{h}{df/dt}\frac{d^2f}{dt^2}\\
        &=\frac{12 G}{c^4 d_L(z)} |P_{\rm DF}| \left(\frac{11}{3} - \frac{d \ln \rho}{d \ln f} - \frac{d \ln C_\Lambda}{d \ln f}\right)~.
\end{align}
The $D$-function is therefore directly related to the DF power and, through the dependence of $P_{\rm DF}$ and $C_\Lambda$ on the properties of the dark environment, contains information about the local DM density and sound speed. It can consequently be used to reconstruct the EoS of the DM.

\bibliographystyle{utphys}
\bibliography{GW_Probe}

\end{document}